\documentclass[preprint1,trackchanges]{aastex62}
\NoNewPageAfterKeywords
\pdfoutput=1 
\usepackage[T1]{fontenc}
\usepackage{apjfonts}
\usepackage[figure,figure*]{hypcap}
\usepackage{makecell}

\newcommand{\dmag}{\Delta_\mathrm{mag}}
\newcommand{\smag}{\sigma_\mathrm{mag}}
\newcommand{\sqdeg}{deg$^2$}
\newcommand{\decd}{$^\circ$}
\newcommand{\rah}{$^\mathrm{h}$}
\newcommand{\ram}{$^\mathrm{m}$}
\newcommand{\ras}{$^\mathrm{s}$}

\shorttitle{OSSOS light curves with HSC}
\shortauthors{Alexandersen et al.}
\watermark{DRAFT}

\begin{document}

\title{OSSOS XII: Variability studies of 65 Trans-Neptunian Objects using the Hyper Suprime-Cam\footnote{Based on data collected at Subaru Telescope, which is operated by the National Astronomical Observatory of Japan.}}

\author[0000-0003-4143-8589]{Mike Alexandersen} 
\correspondingauthor{Mike Alexandersen}
\email{mike.alexandersen@alumni.ubc.ca}
\affiliation{Institute of Astronomy and Astrophysics, Academia Sinica; 11F of AS/NTU Astronomy-Mathematics Building, No. 1 Roosevelt Rd., Sec. 4, Taipei 10617, Taiwan}

\author[0000-0001-8821-5927]{Susan D. Benecchi}
\affiliation{Planetary Science Institute, 1700 East Fort Lowell, Suite 106, Tucson, AZ 85719, USA}

\author[0000-0001-7244-6069]{Ying-Tung Chen}
\affiliation{Institute of Astronomy and Astrophysics, Academia Sinica; 11F of AS/NTU Astronomy-Mathematics Building, No. 1 Roosevelt Rd., Sec. 4, Taipei 10617, Taiwan}

\author[0000-0002-0760-1584]{Marielle R. Eduardo}
\affiliation{Department of Physical Sciences, University of the Philippines Baguio, Gov. Pack Rd., Baguio City, Benguet, Philippines 2600}

\author[0000-0002-1506-4248]{Audrey Thirouin}
\affiliation{Lowell Observatory, 1400 W Mars Hill Rd, Flagstaff, Arizona, 86001, USA}

\author[0000-0003-4365-1455]{Megan E. Schwamb}
\affiliation{Gemini Observatory, Northern Operations Center, 670 North A'ohoku Place, Hilo, HI 96720, USA}

\author[0000-0003-4077-0985]{Matthew J. Lehner}
\affiliation{Institute of Astronomy and Astrophysics, Academia Sinica; 11F of AS/NTU Astronomy-Mathematics Building, No. 1 Roosevelt Rd., Sec. 4, Taipei 10617, Taiwan}
\affiliation{Department of Physics and Astronomy, University of Pennsylvania, 209 S. 33rd St., Philadelphia, PA 19104, USA}
\affiliation{Harvard-Smithsonian Center for Astrophysics, 60 Garden St., Cambridge, MA 02138, USA}

\author[0000-0001-6491-1901]{Shiang-Yu Wang}
\affiliation{Institute of Astronomy and Astrophysics, Academia Sinica; 11F of AS/NTU Astronomy-Mathematics Building, No. 1 Roosevelt Rd., Sec. 4, Taipei 10617, Taiwan}

\author[0000-0003-3257-4490]{Michele T. Bannister}
\affiliation{Astrophysics Research Centre, School of Mathematics and Physics, Queen's University Belfast, Belfast BT7 1NN, United Kingdom}

\author{Brett J. Gladman}
\affiliation{Department of Physics and Astronomy, University of British Columbia, Vancouver, BC V6T 1Z1, Canada}

\author{Stephen D. J. Gwyn}
\affiliation{Herzberg Astronomy and Astrophysics Research Centre, National Research Council of Canada, 5071 West Saanich Rd, Victoria, British Columbia V9E 2E7, Canada}

\author[0000-0001-7032-5255]{JJ. Kavelaars}
\affiliation{Herzberg Astronomy and Astrophysics Research Centre, National Research Council of Canada, 5071 West Saanich Rd, Victoria, British Columbia V9E 2E7, Canada}
\affiliation{Department of Physics and Astronomy, University of Victoria, Elliott Building, 3800 Finnerty Rd, Victoria, BC V8P 5C2, Canada}

\author[0000-0003-0407-2266]{Jean-Marc Petit}
\affiliation{Institut UTINAM UMR6213, CNRS, Univ. Bourgogne Franche-Comt\'e, OSU Theta F25000 Besan\c{c}on, France}

\author[0000-0001-8736-236X]{Kathryn Volk}
\affiliation{Lunar and Planetary Laboratory, University of Arizona, 1629 E University Blvd, Tucson, AZ 85721, USA}

\begin{abstract}

We present variability measurements and partial light curves of Trans-Neptunian Objects (TNOs) from a two-night pilot study using Hyper Suprime-Cam (HSC) on the Subaru Telescope (Maunakea, Hawai'i, USA). 
Subaru's large aperture (8-m) and HSC's large field of view (1.77~\sqdeg) allow us to obtain measurements of multiple objects with a range of magnitudes in each telescope pointing. 
We observed 65 objects with $m_r=22.6$--$25.5$~mag in just six pointings, allowing 20--24~visits of each pointing over the two nights.
Our sample, all discovered in the recent Outer Solar System Origins Survey (OSSOS), span absolute magnitudes $H_r = 6.2$--$10.8$~mag and thus investigates smaller objects than previous light curve projects have typically studied.
Our data supports the existence of a correlation between light curve amplitude and absolute magnitude seen in other works, but does not support a correlation between amplitude and orbital inclination. 
Our sample includes a number of objects from different dynamical populations within the trans-Neptunian region, but we do not find any relationship between variability and dynamical class. 
We were only able to estimate periods for 12 objects in the sample and found that a longer baseline of observations is required for reliable period analysis.
We find that 31 objects (just under half of our sample) have variability $\dmag$ greater than 0.4~mag during all of the observations; in smaller 1.25~hr, 1.85~hr and 2.45~hr windows, the median $\dmag$ is 0.13, 0.16 and 0.19~mags, respectively. 
The fact that variability on this scale is common for small TNOs has important implications for discovery surveys (such as OSSOS or the Large Synoptic Survey Telescope) and color measurements. 
\end{abstract}

\section{Introduction}\label{sec:introduction}

Over the years, rotational light curves (the brightness variation versus time) from small bodies in our Solar System have been used to derive physical properties about the shape, surface, cohesion, internal structure, and density of objects, and to infer binarity \citep{sheppard08, thirouin10, benecchisheppard13}.
Small light curve amplitudes can be caused by albedo variation across the surface of a spherical object, but moderate/large amplitude light curves (maximum minus minimum, $\dmag\geq$0.15-0.2~mag) are likely due to the shape of the object \citep{sheppard08,thirouin10}. 
Flat light curves can be explained by a nearly-spherical object with a homogeneous surface or an elongated object with a pole-on orientation. 
A flat light curve can also be due to an extremely slow rotational period undetectable during the observing window, and such a slow rotation may infer the presence of a companion \citep{thirouin10, benecchisheppard13, thirouin14}. 
Alternatively, very large amplitude light curves ($\dmag\geq$0.9~mag) with a U-/V-shape at the maximum/minimum of brightness are due to close/contact binary systems \citep{sheppardjewitt04, lacerda14a, thirouin17a, thirouin17b, thirouinsheppard18, ryan17}.  

In the trans-Neptunian belt, four main dynamical populations - \emph{classical}, \emph{resonant}, \emph{detached} and \emph{scattering} objects - have been identified with additional sub-divisions inside some of these broad groupings. 
Resonant objects are trapped in mean-motion resonance with Neptune, such that the orbital period (or mean-motion) of the TNO and Neptune have a small integer ratio, which causes the TNO to experience stabilizing gravitational perturbations. 
Objects that are currently interacting strongly with Neptune in such a way that their orbits are unstable are called scattering \citep[a semi-major axis change of at least 1.5~AU in 10~Myr is required for this classification in the nomenclature of][]{gladman08}.
The detached objects are those objects with perihelia that are so distant that they are beyond significant gravitational influence from Neptune; how they got there is a subject of active research.
The resonant, detached and scattering objects are all dynamically excited (``hot'') populations, with wide inclination and eccentricity distributions, thought to have been caused by significant interaction with the giant planets or other small bodies during the era of planetary migration \citep{kaulanewman92,malhotra93,malhotra95,thommes02,kortenkamp04,hahnmalhotra05,murrayclay05,tsiganis05,levison08}.
The classical objects consists of non-resonant, non-scattering objects, primarily located at semi-major axes between 42~AU and 48~AU. 
The classical population has both a hot sub-population and a dynamically cold sub-population of object that have nearly circular orbits with low inclinations and eccentricities; the dynamically cold objects likely had only limited interactions with the giant planets. 
Due to their different formation mechanisms, it is possible that these dynamical populations have different light curve properties, just as they have different color distributions \citep{trujillobrown02, peixinho03,fraserbrown12,sheppard12,tegler16,pike17}. 

Because light curves of TNOs have been studied by several surveys over the past decades, it is now possible to study the rotational properties of the different dynamical populations. 
Several studies have identified different correlations/anti-correlations between orbital and physical/rotational properties, allowing us to understand the different dynamical environments in the trans-Neptunian belt \citep{duffard09,benecchisheppard13,thirouin16}. 
However, most of the past surveys focused on bright objects, as they are the easiest ones to study with small/moderate aperture facilities. 
Therefore, we present a pilot study using the Subaru/HSC to study the rotational properties of the faint TNOs ($H_r>8.0$~mag).
We aim to explore the rotational properties of TNOs in several size ranges, particularly for smaller objects around the break in each population's size distribution \citep{shankman13, fraser14,alexandersen16,lawler18}; however, firm conclusions will require additional observations. 

\section{Sample \& observations}

\begin{figure}[htb]
\includegraphics[width=1.0\textwidth]{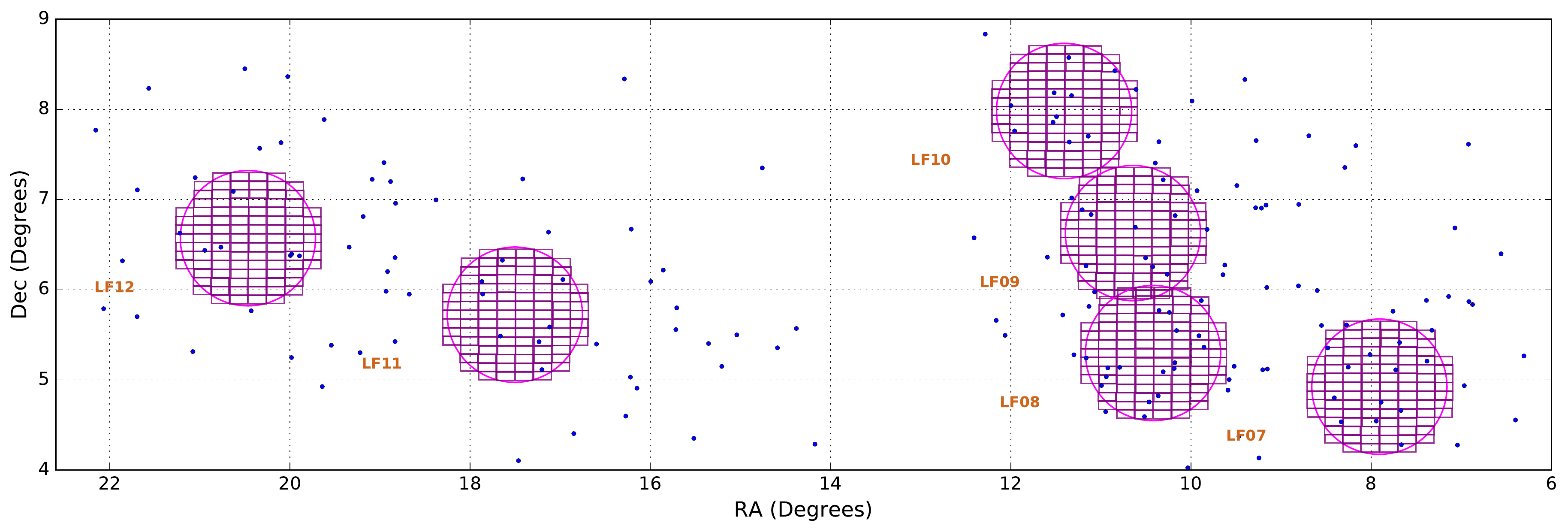}
\caption{%
The location of our 2016 fields relative to TNOs discovered in the OSSOS \texttt{L} and \texttt{S} blocks (left and right, respectively). 
The large field of view of HSC (magenta) allowed us to observe many objects at once. 
The field centers used are listed in \autoref{tab:pointings}. \label{fig:fields}
}
\end{figure}

Our sample consists entirely of objects discovered or detected in the Outer Solar System Origins Survey \citep[OSSOS,][]{bannister16, bannister18}. 
OSSOS was a large program on the Canada-France-Hawaii Telescope (CFHT), which ran from 2013 through 2017.
OSSOS detected and tracked more than 800 trans-Neptunian objects (TNOs) with regular follow-up observations during at least two years to determine accurate orbits and secure dynamical classifications \citep{bannister18}.
Due to the design and success of OSSOS, we have very complete information on the orbital characteristics and classification of these objects.
Likewise, because these objects were discovered recently, they are still tightly clustered on the sky, having only slightly diffused apart due to Kepler shear. 
This allows us to observe up to 17 objects at once within the 1.77~\sqdeg\ field of view of Subaru Telescope's Hyper Suprime-Cam \citep[HSC,][]{miyazaki18}.
The six fields used in this study (coordinates in \autoref{tab:pointings}) were chosen in order to maximize the number of objects within each field (see \autoref{fig:fields}), for a total of 63 TNOs (37 classical, 15 resonant, 5 detached, 6 scattering) and two Centaurs spanning apparent magnitudes $m_r=22.6$--$25.5$~mag and absolute magnitudes $H_r=6.2$--$10.8$~mag (as calculated from our light curve observations). 
This sample far outnumbers (by a factor of 2.5) previously reported variability studies for faint ($H_r\geq 8$~mag) TNOs. 
Our sample thus much more thoroughly probes a smaller size regime than previous light curve studies have typically studied, thanks to the combined aperture size and field of view of Subaru/HSC. 
Information about the individual objects in our sample can be found in \autoref{tab:objects}. 

\begin{deluxetable}{lcccc}[tbp]
\tablecaption{\label{tab:pointings} Field centers used in this work.
On the second night, each field lost one or two observations due to low transparency.}
\tablehead{\colhead{Field} & \colhead{Right Ascension} &  \colhead{Declination} & \colhead{Observations night 1} & \colhead{Observations night 2} \\
\colhead{Identifier} &  \colhead{(hrs)} & \colhead{(deg)} & \colhead{Useful/Total} & \colhead{Useful/Total}}
\startdata
LF07  &  00\rah31\ram33\ras & +04\decd54'54'' & 12/12 & 10/12 \\
LF08  &  00\rah41\ram41\ras & +05\decd17'52'' & 12/12 & 10/12 \\
LF09  &  00\rah42\ram34\ras & +06\decd37'44'' & 12/12 & 10/12 \\
LF10  &  00\rah45\ram37\ras & +07\decd59'02'' & 12/12 & 10/12 \\
LF11  &  01\rah10\ram00\ras & +05\decd43'25'' & 10/10 & 9/10 \\
LF12  &  01\rah21\ram51\ras & +06\decd34'20'' & 10/10 & 9/10 \\
\enddata
\end{deluxetable}

\startlongtable
\begin{deluxetable}{llCCRllRRRR}[p]
\tabletypesize{\footnotesize}
\tablecaption{\label{tab:objects}
Summary of information about the objects in our sample.
\emph{ID} refers to the internal OSSOS designation (as used in \citep{bannister18}).
\emph{Field} is the field in which the target was observed (\autoref{tab:pointings}). 
$a$, $e$ and $i$ are the barycentric orbital parameters semi-major axis, eccentricity and inclination (uncertainties on eccentricity and inclination are always 0.001 or smaller and are therefore not listed here). 
\emph{Class} is the dynamical classification of the object. 
\emph{MPC} is the Minor Planet Center\footnote{https://minorplanetcenter.net/iau/mpc.html} designation (Note: all objects will have MPC designations before final publication.).
$m_r$ and $H_r$ are the mean apparent and absolute magnitudes of the objects, as measured in this work (these are the magnitudes used everywhere in this paper, as they are more accurate than the discovery magnitudes from OSSOS); due to the quality of the OSSOS orbits, the uncertainty on $H_r$ is identical to that of $m_r$ to the given precision.
$\dmag$ is the maximum magnitude variation measured for the object within this work.
$\smag$ is the standard deviation of the magnitudes measured in this work.
This table is ordered in the same order as Figures \ref{fig:hsc_lc1} to \ref{fig:hsc_lc4}: brightest object at the top to faintest object at the bottom.}
\tablehead{%
\colhead{ID} & \colhead{Field} & \colhead{$a$ (au)} & \colhead{$e$} & \colhead{$i$ (deg)} & \colhead{Class} & \colhead{MPC} & \colhead{$m_r$ (mag)} & \colhead{$H_r$ (mag)} & \colhead{$\dmag$} & \colhead{$\smag$}
}
\startdata%
o5t09PD & LF09 &  67.760\pm  0.009  & 0.459 &  3.575 & det       & 2014 UA$_{225}$ & 22.627\pm0.002 &  7.033 & 0.110_{-0.012}^{+0.012} & 0.027_{-0.002}^{+0.002}  \\
o3l43   & LF11 &  45.785\pm  0.004  & 0.097 &  2.025 & cla       & 2013 UL$_{15}$  & 22.989\pm0.002 &  6.702 & 0.363_{-0.011}^{+0.012} & 0.109_{-0.002}^{+0.002}  \\
o5s21   & LF10 &  43.851\pm  0.007  & 0.170 &  2.373 & cla       &                 & 23.225\pm0.003 &  7.564 & 0.36 _{-0.02 }^{+0.02 } & 0.106_{-0.003}^{+0.003}  \\
o3l57   & LF12 &  45.044\pm  0.004  & 0.075 &  1.841 & res 11:6  & 2013 UM$_{15}$  & 23.297\pm0.003 &  6.872 & 0.127_{-0.014}^{+0.015} & 0.034_{-0.003}^{+0.003}  \\
o5t03   & LF09 &  25.967\pm  0.0014 & 0.288 & 18.849 & cen       &                 & 23.380\pm0.003 & 10.825 & 0.15 _{-0.02 }^{+0.03 } & 0.041_{-0.004}^{+0.004}  \\
o5t34PD & LF07 &  45.959\pm  0.003  & 0.170 &  5.081 & res 17:9  & 2003 SP$_{317}$ & 23.511\pm0.006 &  7.226 & 0.56 _{-0.04 }^{+0.05 } & 0.177_{-0.006}^{+0.007}  \\
o5s07   & LF08 &  39.332\pm  0.004  & 0.249 & 16.270 & res 3:2   &                 & 23.615\pm0.004 &  8.957 & 0.22 _{-0.02 }^{+0.02 } & 0.048_{-0.003}^{+0.004}  \\
o5s52   & LF10 &  50.85 \pm  0.03   & 0.247 & 27.196 & res 11:5  &                 & 23.616\pm0.003 &  7.054 & 0.314_{-0.014}^{+0.014} & 0.094_{-0.003}^{+0.003}  \\
o5s36   & LF09 &  43.258\pm  0.003  & 0.053 &  2.206 & cla       &                 & 23.638\pm0.003 &  7.569 & 0.43 _{-0.02 }^{+0.02 } & 0.128_{-0.004}^{+0.004}  \\
o5s05   & LF10 &  21.981\pm  0.010  & 0.479 & 15.389 & cen       & 2015 RV$_{245}$ & 23.640\pm0.004 & 10.767 & 0.10 _{-0.02 }^{+0.02 } & 0.025_{-0.004}^{+0.004}  \\
o3l64   & LF11 &  44.355\pm  0.003  & 0.073 &  1.972 & cla       & 2013 UW$_{16}$  & 23.729\pm0.004 &  7.233 & 0.13 _{-0.02 }^{+0.02 } & 0.038_{-0.004}^{+0.004}  \\
o3l46   & LF12 &  46.612\pm  0.004  & 0.079 &  2.471 & cla       & 2013 UP$_{15}$  & 23.730\pm0.004 &  7.404 & 0.29 _{-0.02 }^{+0.02 } & 0.089_{-0.004}^{+0.004}  \\
o5t08   & LF08 &  38.848\pm  0.003  & 0.057 &  1.097 & cla       &                 & 23.751\pm0.004 &  8.164 & 0.21 _{-0.02 }^{+0.02 } & 0.057_{-0.004}^{+0.004}  \\
o5t30   & LF08 &  43.791\pm  0.008  & 0.073 &  2.170 & cla       &                 & 23.811\pm0.004 &  7.648 & 0.55 _{-0.02 }^{+0.02 } & 0.159_{-0.005}^{+0.005}  \\
o5t42   & LF07 &  39.451\pm  0.011  & 0.204 &  1.587 & res 3:2   &                 & 23.833\pm0.004 &  7.254 & 0.16 _{-0.02 }^{+0.02 } & 0.045_{-0.004}^{+0.004}  \\
o5s14   & LF07 &  34.841\pm  0.002  & 0.047 &  7.122 & res 5:4   &                 & 23.911\pm0.004 &  8.488 & 0.22 _{-0.02 }^{+0.02 } & 0.054_{-0.004}^{+0.004}  \\
o5t51   & LF08 &  59.83 \pm  0.10   & 0.425 & 13.851 & res 14:5  &                 & 23.936\pm0.004 &  6.833 & 0.24 _{-0.02 }^{+0.03 } & 0.057_{-0.004}^{+0.004}  \\
o3l82   & LF11 &  43.657\pm  0.005  & 0.286 & 24.987 & res 7:4   & 2013 UK$_{17}$  & 23.936\pm0.005 &  6.621 & 0.15 _{-0.02 }^{+0.02 } & 0.044_{-0.005}^{+0.006}  \\
o3l63   & LF12 &  45.132\pm  0.004  & 0.054 &  3.363 & cla       & 2013 UN$_{15}$  & 23.986\pm0.005 &  7.493 & 0.56 _{-0.03 }^{+0.03 } & 0.161_{-0.004}^{+0.005}  \\
o3l83   & LF11 & 200.2  \pm  0.5    & 0.781 & 10.652 & det       & 2013 UT$_{15}$  & 24.001\pm0.005 &  6.168 & 0.33 _{-0.02 }^{+0.02 } & 0.098_{-0.005}^{+0.005}  \\
o5s35   & LF09 &  46.40 \pm  0.02   & 0.161 & 24.307 & cla       &                 & 24.065\pm0.005 &  8.023 & 0.37 _{-0.03 }^{+0.03 } & 0.106_{-0.005}^{+0.005}  \\
o3l55   & LF11 &  44.004\pm  0.004  & 0.096 &  2.358 & cla       & 2013 UY$_{16}$  & 24.086\pm0.005 &  7.667 & 0.37 _{-0.02 }^{+0.02 } & 0.121_{-0.005}^{+0.005}  \\
o5t16   & LF08 &  44.79 \pm  0.02   & 0.176 & 17.756 & cla       &                 & 24.158\pm0.005 &  8.358 & 0.36 _{-0.03 }^{+0.04 } & 0.093_{-0.006}^{+0.006}  \\
o5s30   & LF07 &  41.477\pm  0.005  & 0.073 & 33.327 & cla       &                 & 24.164\pm0.007 &  8.247 & 0.27 _{-0.03 }^{+0.03 } & 0.063_{-0.006}^{+0.006}  \\
o3l71   & LF12 &  45.319\pm  0.003  & 0.014 &  1.927 & cla       & 2013 UW$_{17}$  & 24.178\pm0.006 &  7.612 & 0.42 _{-0.03 }^{+0.03 } & 0.125_{-0.005}^{+0.006}  \\
o5t06   & LF08 &  72.06 \pm  0.03   & 0.535 & 12.327 & sca       &                 & 24.183\pm0.006 &  8.941 & 0.25 _{-0.03 }^{+0.03 } & 0.069_{-0.006}^{+0.006}  \\
o5s22   & LF10 &  46.162\pm  0.010  & 0.222 &  8.482 & res 19:10 &                 & 24.207\pm0.005 &  8.538 & 0.21 _{-0.02 }^{+0.02 } & 0.059_{-0.005}^{+0.005}  \\
o5s15   & LF07 &  45.439\pm  0.009  & 0.238 & 25.362 & res 13:7  &                 & 24.246\pm0.006 &  8.809 & 0.31 _{-0.03 }^{+0.03 } & 0.084_{-0.006}^{+0.006}  \\
o3l65   & LF11 &  44.609\pm  0.007  & 0.278 & 11.207 & sca       & 2013 UZ$_{16}$  & 24.272\pm0.006 &  7.775 & 0.36 _{-0.03 }^{+0.03 } & 0.095_{-0.007}^{+0.007}  \\
o3l20   & LF12 &  39.425\pm  0.010  & 0.181 &  2.085 & res 3:2   & 2013 UV$_{17}$  & 24.335\pm0.008 &  8.354 & 0.69 _{-0.04 }^{+0.04 } & 0.185_{-0.008}^{+0.008}  \\
o5s19   & LF10 &  45.054\pm  0.006  & 0.194 & 17.960 & res 11:6  &                 & 24.357\pm0.006 &  8.758 & 0.38 _{-0.03 }^{+0.03 } & 0.105_{-0.006}^{+0.006}  \\
o5t10   & LF07 &  37.844\pm  0.006  & 0.051 &  1.540 & cla       &                 & 24.375\pm0.007 &  8.768 & 0.31 _{-0.04 }^{+0.05 } & 0.081_{-0.008}^{+0.008}  \\
o5t44   & LF08 &  43.560\pm  0.004  & 0.071 &  3.566 & cla       &                 & 24.383\pm0.007 &  7.768 & 0.28 _{-0.03 }^{+0.03 } & 0.085_{-0.006}^{+0.005}  \\
o3l62   & LF11 &  39.230\pm  0.004  & 0.238 &  4.416 & res 3:2   & 2013 UX$_{16}$  & 24.415\pm0.008 &  7.950 & 0.27 _{-0.04 }^{+0.05 } & 0.076_{-0.010}^{+0.010}  \\
o5s24   & LF10 &  50.376\pm  0.018  & 0.278 &  9.376 & res 13:6  &                 & 24.435\pm0.007 &  8.657 & 0.58 _{-0.04 }^{+0.04 } & 0.153_{-0.007}^{+0.007}  \\
o3l19   & LF11 &  45.334\pm  0.003  & 0.126 &  1.480 & cla       & 2013 SM$_{100}$ & 24.447\pm0.007 &  8.507 & 0.68 _{-0.04 }^{+0.04 } & 0.174_{-0.007}^{+0.007}  \\
o5s42   & LF10 &  39.323\pm  0.011  & 0.165 &  5.772 & res 3:2   &                 & 24.482\pm0.007 &  8.284 & 0.38 _{-0.04 }^{+0.05 } & 0.102_{-0.010}^{+0.009}  \\
o5t27   & LF08 &  43.907\pm  0.011  & 0.079 &  1.498 & cla       &                 & 24.490\pm0.007 &  8.348 & 0.42 _{-0.04 }^{+0.04 } & 0.101_{-0.006}^{+0.006}  \\
o5t48   & LF07 &  42.183\pm  0.003  & 0.156 &  5.209 & res 5:3   &                 & 24.526\pm0.011 &  7.693 & 0.41 _{-0.04 }^{+0.04 } & 0.107_{-0.009}^{+0.009}  \\
o5t13   & LF08 &  79.06 \pm  0.03   & 0.535 & 17.629 & res 17:4  &                 & 24.531\pm0.007 &  8.849 & 0.41 _{-0.05 }^{+0.05 } & 0.100_{-0.008}^{+0.008}  \\
o5t32   & LF08 &  43.791\pm  0.005  & 0.035 &  2.209 & cla       &                 & 24.537\pm0.008 &  8.292 & 0.48 _{-0.05 }^{+0.05 } & 0.117_{-0.010}^{+0.010}  \\
uo3l84  & LF11 &  44.012\pm  0.003  & 0.082 &  1.435 & cla       &                 & 24.547\pm0.009 &  7.858 & 0.44 _{-0.06 }^{+0.06 } & 0.117_{-0.010}^{+0.011}  \\
o3l44   & LF12 &  42.793\pm  0.005  & 0.042 &  2.940 & cla       & 2013 UC$_{18}$  & 24.557\pm0.008 &  8.261 & 0.47 _{-0.04 }^{+0.04 } & 0.141_{-0.009}^{+0.010}  \\
uo3l88  & LF11 &  43.935\pm  0.004  & 0.081 &  2.479 & cla       &                 & 24.558\pm0.007 &  8.275 & 0.53 _{-0.05 }^{+0.04 } & 0.169_{-0.008}^{+0.009}  \\
o5t07   & LF07 &  39.513\pm  0.009  & 0.192 &  2.802 & res 3:2   &                 & 24.611\pm0.014 &  9.159 & 0.48 _{-0.06 }^{+0.06 } & 0.143_{-0.014}^{+0.015}  \\
o5t41   & LF08 &  44.310\pm  0.004  & 0.043 &  0.997 & cla       &                 & 24.659\pm0.011 &  8.085 & 0.23 _{-0.04 }^{+0.05 } & 0.068_{-0.009}^{+0.012}  \\
o5t22   & LF08 &  39.58 \pm  0.02   & 0.176 &  5.457 & res 3:2   &                 & 24.664\pm0.009 &  8.612 & 0.28 _{-0.03 }^{+0.05 } & 0.076_{-0.008}^{+0.009}  \\
o5t49   & LF08 &  44.851\pm  0.004  & 0.096 &  1.968 & cla       &                 & 24.672\pm0.008 &  7.807 & 0.64 _{-0.04 }^{+0.04 } & 0.204_{-0.008}^{+0.008}  \\
o5s61   & LF10 &  44.942\pm  0.012  & 0.117 &  3.023 & cla       &                 & 24.696\pm0.009 &  7.879 & 0.65 _{-0.04 }^{+0.05 } & 0.177_{-0.009}^{+0.009}  \\
o5t23   & LF08 &  46.215\pm  0.007  & 0.132 &  5.089 & cla       &                 & 24.813\pm0.009 &  8.760 & 0.66 _{-0.06 }^{+0.08 } & 0.174_{-0.012}^{+0.012}  \\
o5s40   & LF09 &  44.310\pm  0.007  & 0.075 &  2.276 & cla       &                 & 24.965\pm0.010 &  8.702 & 0.41 _{-0.04 }^{+0.05 } & 0.117_{-0.009}^{+0.009}  \\
o5t26   & LF09 &  43.173\pm  0.004  & 0.043 & 16.024 & cla       &                 & 24.973\pm0.013 &  8.855 & 0.41 _{-0.05 }^{+0.07 } & 0.115_{-0.014}^{+0.014}  \\
uo5t55  & LF08 &  44.25 \pm  0.02   & 0.073 &  1.544 & cla       &                 & 25.015\pm0.011 &  8.702 & 0.54 _{-0.07 }^{+0.08 } & 0.129_{-0.013}^{+0.013}  \\
o5s29   & LF07 &  41.396\pm  0.003  & 0.045 & 34.929 & cla       &                 & 25.023\pm0.012 &  9.111 & 0.33 _{-0.04 }^{+0.05 } & 0.097_{-0.011}^{+0.011}  \\
o5s43   & LF07 &  70.84 \pm  0.07   & 0.465 & 10.192 & det       &                 & 25.044\pm0.011 &  8.837 & 0.49 _{-0.08 }^{+0.07 } & 0.120_{-0.015}^{+0.014}  \\
o5t25   & LF08 &  43.288\pm  0.006  & 0.047 &  0.461 & cla       &                 & 25.084\pm0.012 &  8.971 & 0.40 _{-0.06 }^{+0.06 } & 0.099_{-0.011}^{+0.011}  \\
o5t21   & LF08 &  43.633\pm  0.003  & 0.074 &  6.872 & res 7:4   &                 & 25.104\pm0.014 &  9.082 & 0.59 _{-0.06 }^{+0.07 } & 0.165_{-0.013}^{+0.013}  \\
o5s55   & LF09 &  46.93 \pm  0.02   & 0.173 &  4.812 & cla       &                 & 25.107\pm0.010 &  8.528 & 0.59 _{-0.06 }^{+0.07 } & 0.148_{-0.010}^{+0.011}  \\
o5s26   & LF09 &  41.344\pm  0.007  & 0.112 & 17.483 & cla       &                 & 25.111\pm0.014 &  9.299 & 0.54 _{-0.05 }^{+0.05 } & 0.152_{-0.012}^{+0.012}  \\
o5s12   & LF10 &  34.791\pm  0.005  & 0.100 &  2.668 & res 5:4   &                 & 25.118\pm0.015 &  9.798 & 0.40 _{-0.06 }^{+0.08 } & 0.13 _{-0.02 }^{+0.02 }  \\
o5s49   & LF07 &  44.017\pm  0.007  & 0.044 &  2.088 & cla       &                 & 25.178\pm0.012 &  8.889 & 1.01 _{-0.09 }^{+0.10 } & 0.22 _{-0.02 }^{+0.02 }  \\
o5s60   & LF10 &  42.530\pm  0.007  & 0.128 & 40.362 & cla       &                 & 25.180\pm0.012 &  8.510 & 0.51 _{-0.06 }^{+0.07 } & 0.127_{-0.012}^{+0.012}  \\
o5t24   & LF09 &  44.953\pm  0.006  & 0.107 &  2.398 & cla       &                 & 25.257\pm0.014 &  9.199 & 0.80 _{-0.08 }^{+0.08 } & 0.20 _{-0.02 }^{+0.02 }  \\
o5s57   & LF07 &  41.356\pm  0.011  & 0.140 & 32.358 & cla       &                 & 25.276\pm0.016 &  8.649 & 0.84 _{-0.09 }^{+0.10 } & 0.227_{-0.015}^{+0.016}  \\
uo5s77  & LF09 &  65.95 \pm  0.11   & 0.413 & 38.607 & res 13:4  &                 & 25.604\pm0.018 &  9.477 & 0.98 _{-0.09 }^{+0.09 } & 0.295_{-0.02 }^{+0.02 }  \\
\enddata%
\tabletypesize{\normalsize}
\end{deluxetable}

Data were collected by M.~Alexandersen and S.-Y.~Wang on the nights of 2016 August 25 and 26, using HSC on Subaru Telescope, located on Maunakea, Hawai'i, USA\footnote{All data obtained as part of this project can be downloaded from https://smoka.nao.ac.jp/fssearch or http://www.cadc-ccda.hia-iha.nrc-cnrc.gc.ca/en/search/}.
\autoref{fig:fields} shows our field locations in relation to the swarm of OSSOS TNOs. 
As TNOs are typically brightest in r-band, this filter was used. 
The six fields were observed in a repeated sequence with one 300~second exposure of each field, resulting in about 36~minutes between repeat observations of the same field.
Because the observing nights were approximately 1.5 months from opposition, we were unable to observe our fields for the entire night; the four fields closer to opposition thus were observed 12 times each night, while two other fields were observed 10 times each night.
Observations were taken at airmasses ranging from 1.02 to 2.04. The conditions were generally good (transparency near 1.0, seeing mostly between 0.65'' and 1.0''), apart from about one hour of clouds and very low transparency on the second night. One or two  observations of each field were therefore unusable for this project.

\section{Photometry \& calibration}\label{sec:analysis}

Images were processed using \texttt{hscPipe}\footnote{\url{https://hscdata.mtk.nao.ac.jp:4443/hsc_bin_dist/index.html}}, the official HSC data reduction software package, which includes calibration of astrometry/photometry and basic image processing (bias, darks, and flats) \citep{bosch18}.
\texttt{hscPipe} was developed as a prototype pipeline for the Large Synoptic Survey Telescope (LSST) project \citep{ivezic08}. 
The algorithms and conceptions of \texttt{hscPipe} were inherited from the Sloan Digital Sky Survey (SDSS) Photo Pipeline \citep{lupton01}. 
The standard outputs of the \texttt{hscPipe} process include detection catalogs and calibrated images.
We used the Pan-STARRS1 PV2 catalog \citep[PS1,][]{tonry12,schlafly12,magnier13} as the reference catalog for photometry and astrometry.

Photometry of bright point sources revealed systematic variability (of order 0.01~magnitudes), caused by the fact that \texttt{hscPipe} calibrates images individually; images of the same field can thus have slight discrepancies relative to each other due to uncertainties in the zero-point determinations.  
To mitigate for this effect, background stars were inspected to find a set of 10-15 non-variable stars, defined as those that exhibit similar systematic variability, to reveal the systematic variability of the zero-point error. 
A secondary correction was then applied to the zero points, such that the average magnitude of the non-variable stars is constant across the set of images. 
This adjustment was typically smaller than 0.02~magnitudes, but is important for obtaining the most accurate TNO photometry possible.

In order to get the best possible photometric measurements of our TNOs, we used the Python package \texttt{TRIPPy} \citep{fraser16}. 
TNOs move at typical rates of 3--6''/hr at opposition, thus moving up to 0.5'' during our 300-second exposures, causing slight trailing. 
\texttt{TRIPPy} uses knowledge of the rate and angle of motion to accurately measure photometry for moving objects using an elongated aperture and appropriate aperture correction. 

We obtained 9--23 useful measurements for each object over two nights; the median number of measurements per object is 19. 
One or two images per field were discarded due to low transparency during one hour on the second night; while some reduced transparency can be corrected for in calibration, we found that images with transparency $<50\%$ consistently produced anomalous results, so we chose to exclude those images. 
Many objects lost one or more measurement due to being too close to a background source for reliable photometry. 
Three objects (o5t41, o5t07 and o5s12) only have data on one night because they fell into chip gaps on one of the two nights; this was difficult to avoid when optimizing field location for up to 17 objects moving at different rates. 
One object (o5t44) was on a different chip on the two nights; however, we are confident that our zero point calibration is accurate such that there is not an offset between the two nights. 
The final calibrated photometric measurements as well as the calibrated zero points can be found in \autoref{tab:photometry}. 

\startlongtable
\begin{deluxetable}{CCC}[tbp]
\tablecaption{\label{tab:photometry} Photometric measurements from this HSC data set.
$\sigma_\mathrm{Rand}$ is the random uncertainty of the $m_r$ magnitude measurement.
$Zp$ is the zero-point used. 
$\sigma_\mathrm{Sys}$ is the systematic uncertainty on the zero-point (and thus also on the measurement).
This is an example for a single object; all measurements from all 65 objects are available in machine readable format in the on-line supplements. 
}
\tablehead{%
\colhead{MJD$-57620$} & \colhead{$m_r$ $\pm$ $\sigma_\mathrm{Rand}$} & \colhead{$Zp$ $\pm$ $\sigma_\mathrm{Sys}$}
}
\startdata%
          & \textrm{o3l19}     &                     \\
6.3913125 & 24.09   \pm 0.03   & 26.995  \pm 0.005   \\
6.4154337 & 24.21   \pm 0.02   & 27.012  \pm 0.005   \\
6.4393438 & 24.45   \pm 0.02   & 27.022  \pm 0.005   \\
6.4634385 & 24.69   \pm 0.04   & 27.058  \pm 0.005   \\
6.4876346 & 24.78   \pm 0.03   & 27.047  \pm 0.005   \\
6.5123402 & 24.56   \pm 0.03   & 27.039  \pm 0.005   \\
6.5375195 & 24.41   \pm 0.03   & 27.043  \pm 0.005   \\
6.5628085 & 24.29   \pm 0.03   & 27.045  \pm 0.005   \\
6.5875091 & 24.28   \pm 0.02   & 27.038  \pm 0.005   \\
7.3876072 & 24.55   \pm 0.03   & 26.978  \pm 0.005   \\
7.4114040 & 24.67   \pm 0.03   & 27.002  \pm 0.005   \\
7.4353624 & 24.45   \pm 0.02   & 26.885  \pm 0.005   \\
7.4595078 & 24.47   \pm 0.02   & 26.869  \pm 0.005   \\
7.5081300 & 24.36   \pm 0.03   & 26.497  \pm 0.005   \\
7.5584662 & 24.35   \pm 0.02   & 27.019  \pm 0.005   \\
7.5830743 & 24.42   \pm 0.02   & 27.004  \pm 0.005   \\
7.6079518 & 24.58   \pm 0.02   & 26.994  \pm 0.005   \\
\enddata%
\end{deluxetable}

\section{Results}\label{sec:results}

\begin{figure}[htbp]
\plotone{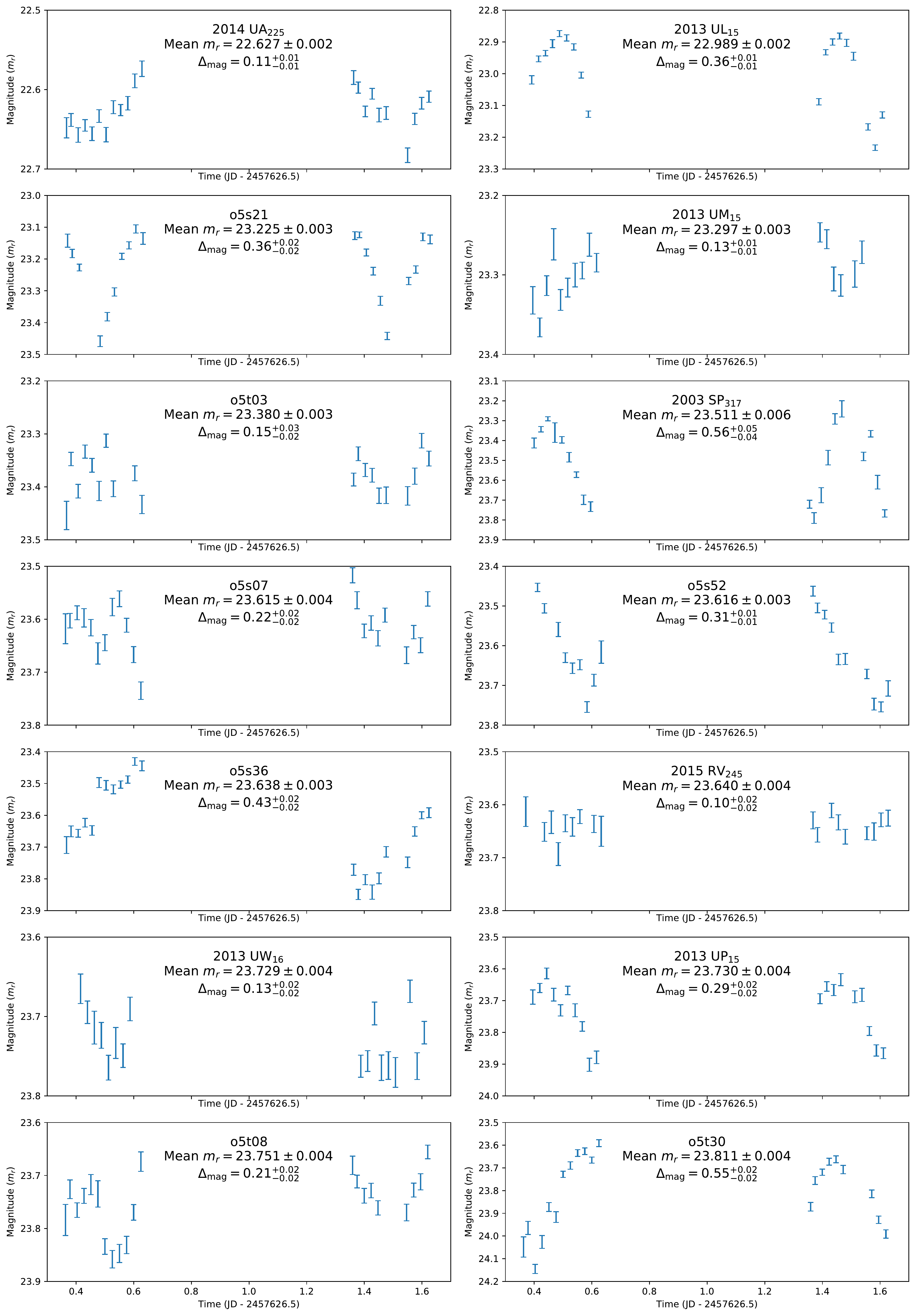}
\caption{\label{fig:hsc_lc0}%
Absolutely calibrated photometry of our TNO sample.
Objects 0--13, ordered by average magnitude. 
}
\end{figure}

\begin{figure}[htbp]
\plotone{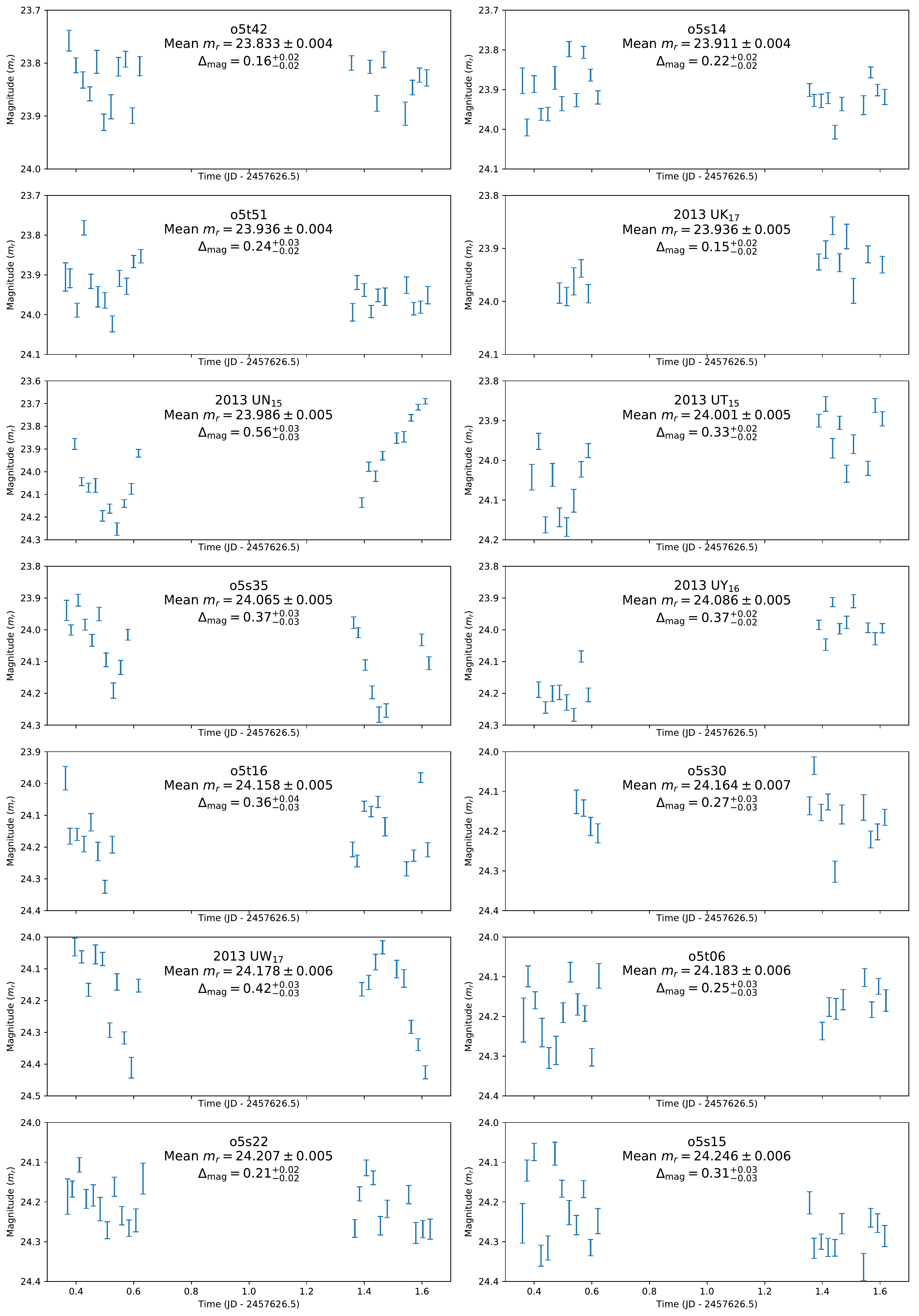}
\caption{\label{fig:hsc_lc1}%
Absolutely calibrated photometry of our TNO sample.
Objects 14--27, ordered by average magnitude. 
}
\end{figure}

\begin{figure}[htbp]
\plotone{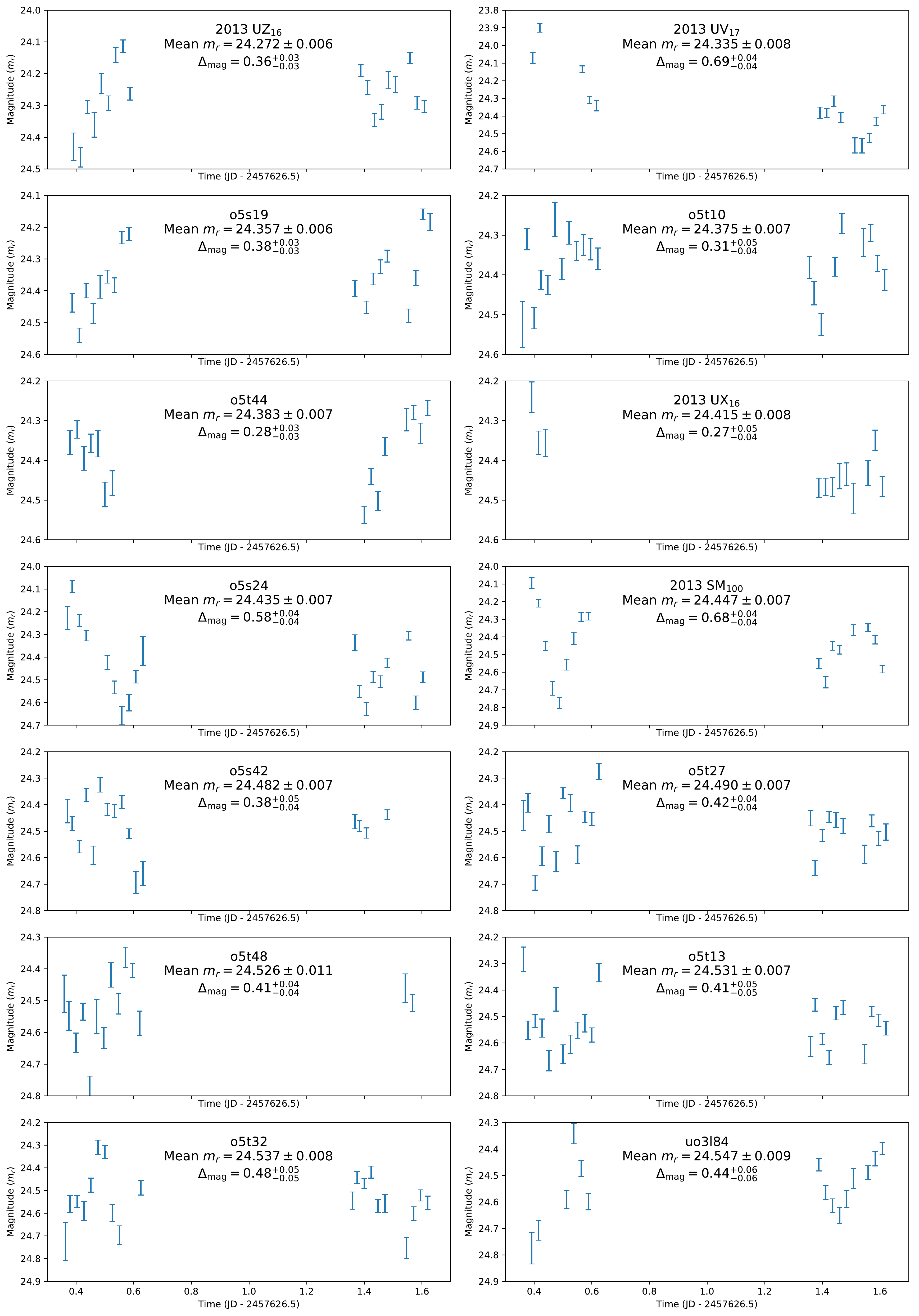}
\caption{\label{fig:hsc_lc2}%
Absolutely calibrated photometry of our TNO sample.
Objects 28--41, ordered by average magnitude. 
}
\end{figure}

\begin{figure}[htbp]
\plotone{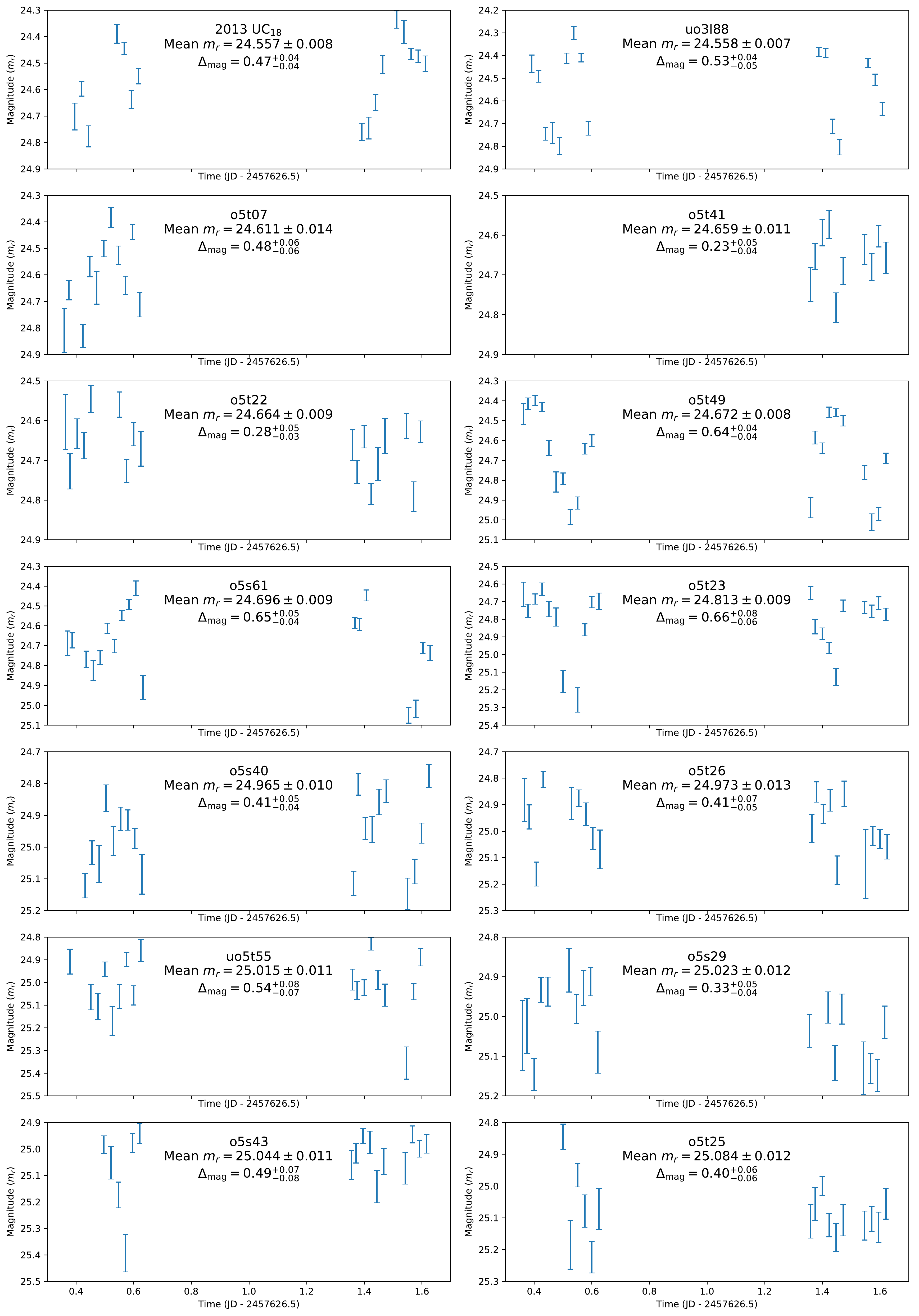}
\caption{\label{fig:hsc_lc3}%
Absolutely calibrated photometry of our TNO sample.
Objects 42--55, ordered by average magnitude. 
}
\end{figure}

\begin{figure}[htbp]
\plotone{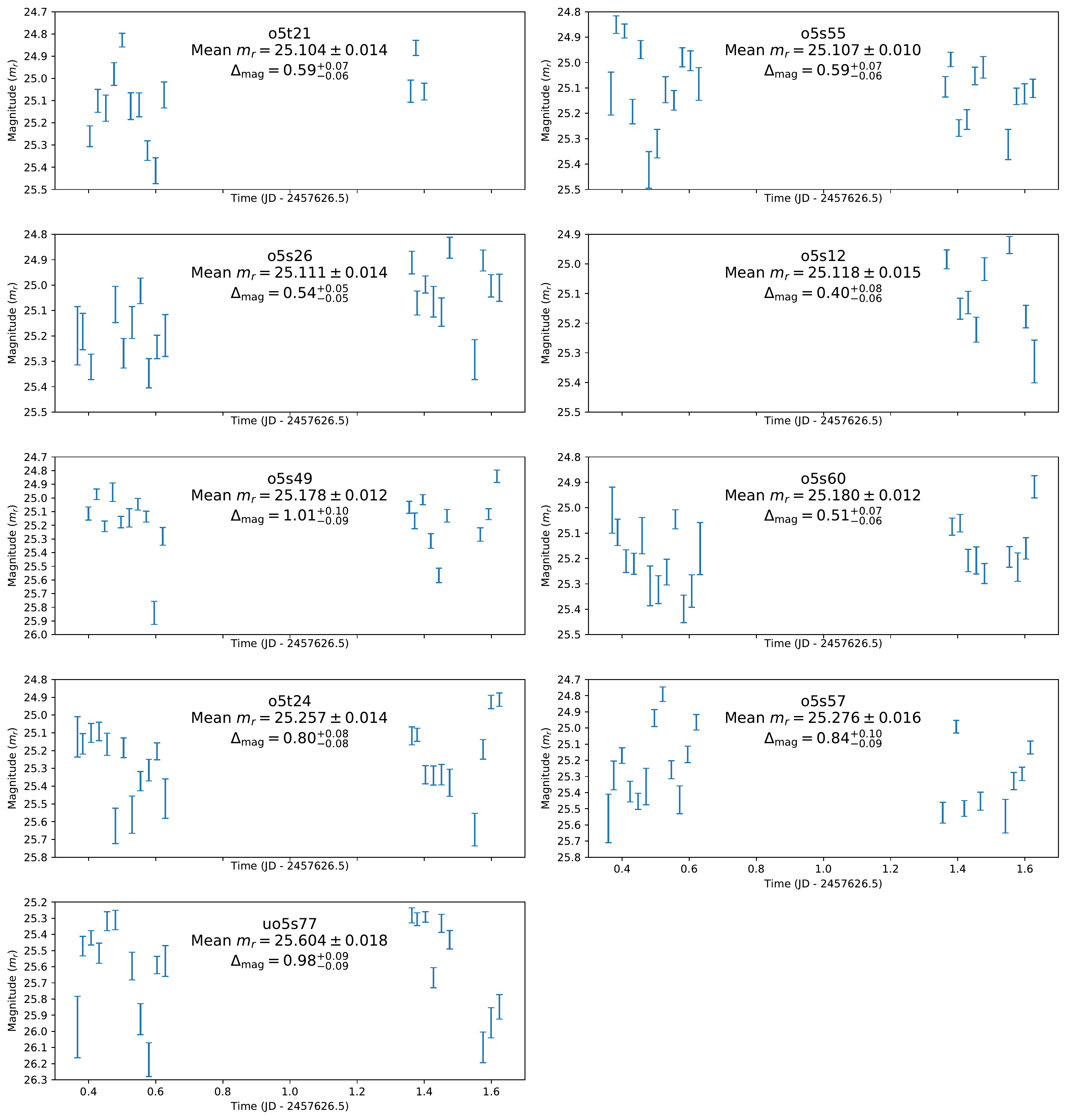}
\caption{\label{fig:hsc_lc4}%
Absolutely calibrated photometry of our TNO sample.
Objects 56--64, ordered by average magnitude. 
}
\end{figure}

We have measured partial light curves for 63 TNOs and two Centaurs. 
The measurements can be found in graphical form in Figures \ref{fig:hsc_lc0} to \ref{fig:hsc_lc4}; the photometric data can be found in \autoref{tab:photometry}.
Several objects (such as 2013 UL$_{15}$, o5s21, 2003 SP$_{317}$, o5t30 and 2013 UN$_{15}$) feature clear, high-amplitude variability.
We have investigated the relationships between our measured light curve variability and other properties such as absolute magnitude, dynamical class, and other orbital parameters (\autoref{sec:amplitude}). 
For the best quality light curves, estimates of the rotation periods were also made (\autoref{sec:period}).  

In order to estimate the uncertainty on parameters such as the magnitude variation, period, correlation coefficient and so on, a resampling method was used.
One thousand resampled datasets were generated by resampling each magnitude measurement of each object. This resampling was drawn from a normal distribution centered on the measured value with a width equal to the measurement uncertainty. 
The analysis was then performed on each resampled dataset independently to find the distribution of resultant values. 
The $2\sigma$ uncertainty on a value is then taken to be the range that spans the central 95\% of values (ie. 2.5\% in each tail). 

\subsection{Amplitude analysis}\label{sec:amplitude}

The mean rotational period of TNOs has been found to be 7--9 hours \citep{duffard09, benecchisheppard13, thirouin16}, although periods range from 3.9 hours (Haumea) to 154 hours (Pluto). 
While we don't have full light curves for all of our target objects, the mean period of 7--9 hours means that we likely observed a near-maximum and near-minimum of most of our target objects. 
Simple simulations of our observations confirm that for objects of a given period between 3 and 40 hours, we observe more than 80\% of the variability on average (averaged over random starting phase); for periods between 3 and 22 hours, we observe more than 88\% of the variability on average. 
It is thus reasonable to assume that our measurements of $\dmag$ (maximum minus minimum magnitude) are close to the full light curve amplitudes and that the slight underestimation will not affect our results (such as whether or not correlations exist).
In this subsection, we have removed the two Centaurs (2015 RV$_{245}$ and o5t03) from the sample, because the surface and rotation of these objects might have been altered by close approaches to the Sun or planets \citep{thirouin10,duffard09}.
As an alternative to the difference between the brightest and faintest observation ($\dmag$), we also examine the standard deviation ($\smag$) of all the observations of each object, because the standard deviation is a measure of overall variation that is less influenced by individual anomalous measurements.

\begin{figure}[tbp]
\includegraphics[width=1.0\textwidth]{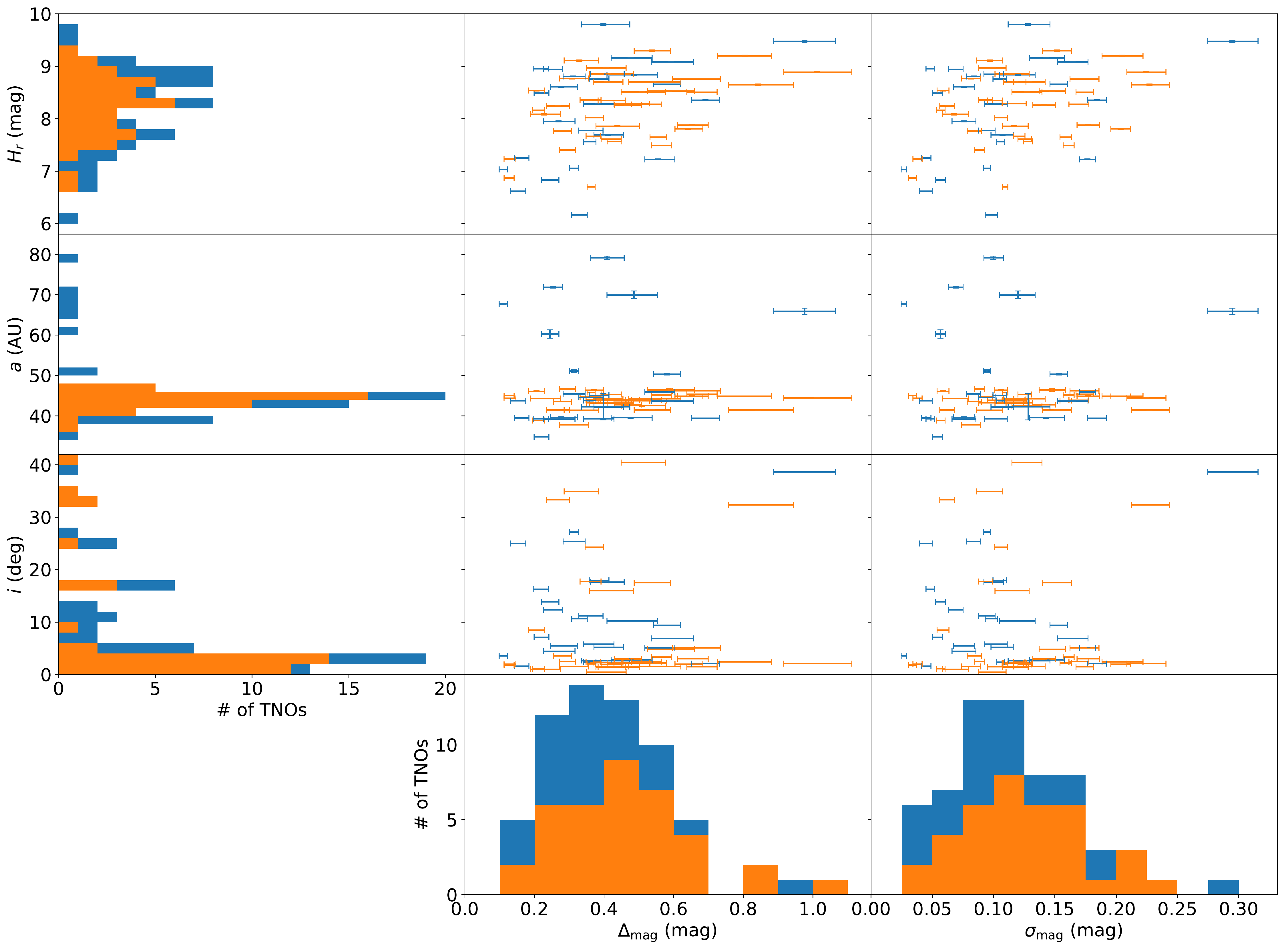}
\caption{\label{fig:hist+scat}%
Left and bottom sides: histograms of $H_r$ magnitude, semi-major axis $a$, inclination $i$, $\dmag$ and $\smag$ of our data.
Center and top right: scatter plots of these parameters. 
Classical TNOs are shown in orange, and non-classical objects are shown in blue.
}
\end{figure}

In \autoref{fig:hist+scat}, we plot both $\dmag$ and $\smag$ versus object $H_r$, semi-major axis $a$, and inclination $i$ to illustrate possible relations between these parameters.
In a sample of 128 objects, \citet{benecchisheppard13} found a statistically significant correlation ($\rho = 0.288$, $P = 0.001$) between light curve amplitude and absolute magnitude. 
Their sample spanned absolute magnitudes from 0 to 12~mag, although most were between 4 and 7~mag; very few had absolute magnitude fainter than 8~mag.
Our sample of 63 TNOs span $H_r=6.2$--$10.8$~mag\footnote{Whenever we discuss apparent and absolute magnitudes for our sample objects, those magnitudes are measured/calculated from our light curves observations, as these are more accurate than the discovery magnitudes reported by OSSOS.}, with more than half the sample at $H_r>8.0$~mag; our sample thus represents variability properties for fainter TNOs than the bulk of the \citet{benecchisheppard13} sample.
Using the Spearman Rank correlation test \citep{spearman1904} and our independent sample of 63 TNOs, we find $\rho = 0.37$, $P = 0.003$ for $\dmag$ versus $H_r$ and $\rho = 0.31$, $P = 0.013$ for $\smag$ versus $H_r$. 
Our data thus strongly support the existence of a weak relationship where intrinsically fainter (likely smaller) bodies on average have larger light curve amplitudes. 
This is likely due to increasingly elongated shapes.
\citet{benecchisheppard13} also found a borderline significant anti-correlation ($\rho = -0.170$, $P = 0.056$) between light curve amplitude and inclination ($i$), however, our data do not support the existence of such an anti-correlation.

Because the cold classicals are thought to have formed in place while all the hot populations (hot classicals, resonant, detached and scattering) are thought to have been scattered outwards to some extent after having formed closer to the Sun \citep[e.g.][]{li08,levison08,batygin11}, we also test the dynamically cold and hot samples separately to identify whether a trend might be present in one population and not in the other. 
In this work, we use the term ``hot population'' to refer to all of the dynamically excited populations combined \citep[as done in][and others]{fraser14}; specifically our hot population sample contains everything non-classical plus classical objects with $i\geq6^\circ$. 
Our ``cold classicals'' sample consists only of classicals with $i\leq4^\circ$. 
Classicals with $4^\circ<i<6^\circ$ (two objects in our sample) are excluded from either sample to minimize contamination. 
The correlation values as well as their significance levels can be found in \autoref{tab:spearman} and are simply summarized here.
For our samples of 26 cold classical TNOs and 35 hot TNOs, we find that both have similar correlation coefficients to the full sample for amplitude versus $H_r$, but with a lower significance than in the full sample (likely due mostly to the now smaller sample sizes). 
For amplitude versus inclination, we find that neither population has a statistically significant correlation, just as in the overall sample.  
Because the semi-major axis ($a$) is possibly related to where objects formed, we also tested for correlations between the light curve variation and $a$ for the full sample and the hot population (the cold population was not tested due to its narrow $a$ range). 
We find that the semi-major axis is not at all correlated with the variability in the full sample nor in the hot population. 

\begin{deluxetable}{CCCCC}[tbp]
\tablecaption{\label{tab:spearman} Spearman Rank correlation test results.
Tests for correlation between the measured light curve variation $\dmag$ and $\smag$ and the parameter listed in the first column.
}
\tablehead{%
\colhead{Parameter} & \colhead{$\rho_{\dmag}$} & \colhead{$P_{\dmag}$}  & \colhead{$\rho_{\smag}$} & \colhead{$P_{\smag}$}
}
\startdata%
                                                     \multicolumn5c{All objects}                                    \\
H_r           &  0.37_{-0.03}^{+0.03} &  0.003 &  0.31_{-0.02}^{+0.03} &  0.013   \\
a             &  0.10_{-0.03}^{+0.03} &  0.43  &  0.11_{-0.02}^{+0.02} &  0.41    \\
i             & -0.05_{-0.03}^{+0.03} &  0.7   & -0.07_{-0.02}^{+0.02}  &  0.60   \\
                                         \multicolumn5c{Cold classicals}                                            \\
H_r           &  0.43_{-0.05}^{+0.04} &  0.029  &  0.30_{-0.04}^{+0.03} &  0.14    \\
i             &  0.25_{-0.05}^{+0.05} &  0.21   &  0.38_{-0.04}^{+0.03} &  0.06    \\
                                         \multicolumn5c{Hot populations}                                            \\
H_r           &  0.42_{-0.05}^{+0.04} &  0.013 &  0.41_{-0.04}^{+0.03} &  0.015  \\
a             &  0.04_{-0.05}^{+0.05} &  0.79  &  0.04_{-0.04}^{+0.04} &  0.79    \\
i             &  0.08_{-0.04}^{+0.05} &  0.6   &  0.03_{-0.03}^{+0.03} &  0.85    \\
\enddata%
\end{deluxetable}

To further test the notion that different populations may have different amplitude distributions, the 2-sample Anderson-Darling test was used to compare subsamples. 
The subsamples used were the classical, resonant, hot and cold populations.
We did not look at the scattering and detached objects in this analysis because the sample sizes are too small for any meaningful results.
In every case, the subsample was tested against the remainder of the sample (ie. classicals versus everything non-classical, etc). 
See \autoref{tab:AD2} for the resultant $P$ values. 
No subsample was found to be significantly different from the rest of the sample. 

\begin{deluxetable}{lCCC}[tbp]
\tablecaption{2-sample Anderson-Darling test results. 
Tests for whether two samples are likely to have been drawn from the same source population.
Here we test whether the light curve variability (measured by $\dmag$ and $\smag$) distribution is different for different dynamical sub-populations.
In every case here, a sub-sample is tested against the remainder of our sample (eg. the resonant objects against everything that is not resonant). \label{tab:AD2}}
\tablehead{%
\colhead{sub-population} & \colhead{$N_\mathrm{obj}$} & \colhead{$P_{\dmag}$} & \colhead{$P_{\smag}$}
}
\startdata%
classicals               & 37 & 0.15_{-0.07}^{+0.10}  & 0.11_{-0.05}^{+0.06} \\
resonant                 & 15 & 0.8\pm0.2             & 0.76_{-0.14}^{+0.12} \\
hot                      & 35 & 0.13_{-0.06}^{+0.09}  & 0.09_{-0.03}^{+0.06} \\
cold                     & 26 & 0.4\pm0.2             & 0.29_{-0.09}^{+0.12}  \\
\enddata%
\end{deluxetable}

\subsection{Period analysis}\label{sec:period}

The rotation periods of TNOs range from 3.9 hours (Haumea) to 154 hours (Pluto), with a mean of 7--9 hours \citep{duffard09, benecchisheppard13, thirouin16}; there are clearly a number of assumptions and biases that go into this mean, not the least of which is the aliasing effects from observations on consecutive nights and the fact that tracking low amplitude and slowly varying objects requires a lot of telescope time. 
Both of these biases exist in our dataset; nevertheless, we seek to use the data we present here to at least place limits on the rotation periods and amplitudes of the objects in our sample.
In the few cases where our data allow for more extensive interpretation, we proceed as appropriate for each object. 
We attempt to fit periods to our light curve data using two methods: a Phase Dispersion Minimization (PDM) method and a Power Spectral Density (PSD) method. 
Because our TNOs are all small enough that we do not expect them to be spherical, they should all have double peaked periods.
The two light curve peaks of an elongated TNO are often difficult to distinguish (for a perfectly symmetrical ellipsoid, the two peaks would be identical).
These methods therefore simply identify the period of the light curve, not the TNO.
We double the single-peaked periods found by our two period methods to get the most likely rotation period of the TNO.

Because our observing window is about 30 hours (from the start of night one to the end of night two) we are able to sample rotation periods from the rotational break-up-limit of 3.3 hours \citep{romanishintegler99} to about 60 hours (where we are sampling half a rotation, which is a full peak-to-peak cycle for a double-peaked light curve). 
For the PSD method, we therefore fold the light curves using every period from 1.0 hours to 30.0 hours in 0.1 hour intervals. 
The Fourier coefficients are calculated from each folded light curve; the sum of the squares of the Fourier coefficients as a function of period is the PSD.
The period that has the highest PSD value is selected as the most likely period of the TNO.
In the PDM method, we fit the data points for each object using a modified PDM \citep{stellingwerf78, buie18}. 
This method goes through every possible period, folds the data and fits a second-order Fourier series to each folded light curve. 
The quality of each fit is captured from the residuals; the fits are ranked and the fit with the highest quality is selected. 

\begin{figure}[tbp]
\includegraphics[width=1.0\textwidth]{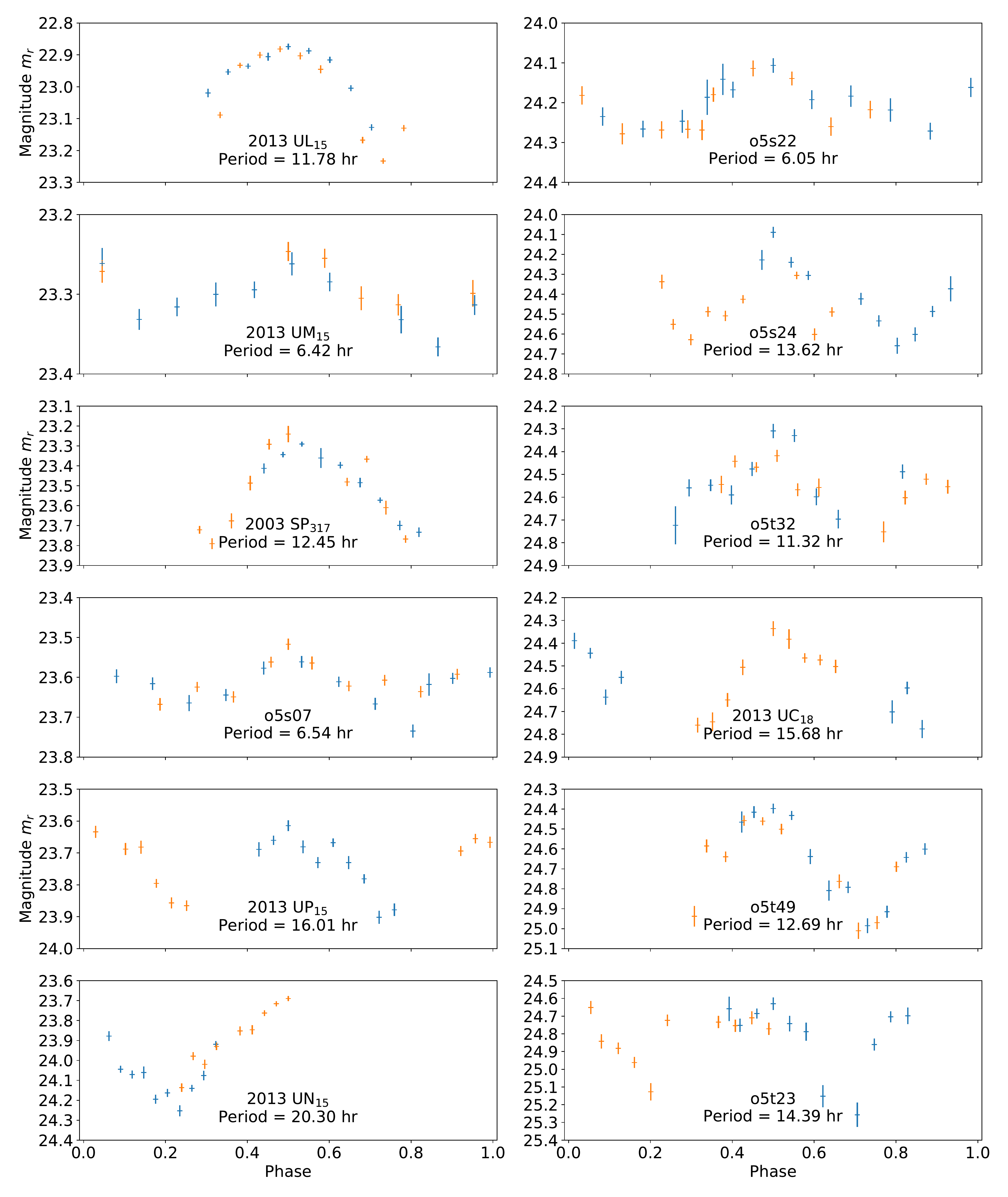}
\caption{\label{fig:folded}%
Folded light curves for the 12 objects for which we have a period estimate.
Blue and orange points show data from the first and second nights, respectively.
The arbitrary phase=0 point was chosen such that the brightest measured magnitude falls at phase=0.5. 
The exact values of MJD, magnitude and uncertainty of all observations can be found in \autoref{tab:photometry}.
}
\end{figure}

\autoref{tab:periods} lists the periods found from these two methods (converted into double-peak periods) for objects where the two methods agree; \autoref{fig:folded} shows the light curves folded at those periods.
When the two methods do not agree, it is likely evidence that we do not have enough data to determine the period with confidence.
The PDM curves from our period analysis is available in the supplementary materials, Figures \ref{fig:pdm0} to \ref{fig:pdm4}.
Although the mean period from our dataset analysis is somewhat greater than that given in the literature ($\sim12.3$ hours compared with 7-9 hours, \citet{duffard09, benecchisheppard13, thirouin10}) it is important to recognize that the periods we are reporting are based on single epoch measurements without the same fidelity of many of the light curves in the literature. 
The literature contains both single and double-peaked light curves and light curves of binary objects. 
We cannot distinguish among these interpretations and have assumed a double-peaked light curve for all objects, so one should not over-interpret this apparent result from our dataset.
This sample of 12 periods is insufficient to draw any conclusions, other than the fact that we would need a larger baseline than two consecutive 6-hour nights to determine the period of a larger fraction of objects.
Adding a third night and observing closer to opposition (when the targets are visible for 8 hours) would allow determination of periods for a much larger fraction of light curves. 
We therefore leave discussion of the period distribution for future work.

\begin{deluxetable}{lRLRL}
\tablecaption{Best periods from two methods, in hours. 
Numbers in parentheses are the percentage of resampled light curves that had a ``best'' period within $5\sigma$ of the reported value. 
Only objects where the two methods agree on the best period are included. 
As these objects are small, their variability is most likely shape-dominated; shape dominated light curves have two peaks per rotation of the object.  
\label{tab:periods}}
\tablehead{%
\colhead{Object} & \colhead{Period$_\mathrm{PDM}$ (hour)} &  & \colhead{Period$_\mathrm{PDS}$ (hour)} &
}
\startdata%
2013 UL$_{15}$   & 11.76\pm 0.02 & ( 41.3\%)   & 11.80\pm 0.03 & (100.0\%)\\
2013 UC$_{18}$   & 15.66\pm 0.23 & ( 80.7\%)   & 15.70\pm 0.08 & ( 68.8\%)\\
2013 UP$_{15}$   & 16.02\pm 0.09 & ( 37.7\%)   & 16.00\pm 0.07 & (100.0\%)\\
2013 UM$_{15}$   &  6.44\pm 0.03 & ( 81.0\%)   &  6.40\pm 0.04 & ( 49.0\%)\\
2013 UN$_{15}$   & 20.30\pm 0.08 & ( 79.2\%)   & 20.30\pm 0.09 & ( 43.6\%)\\
o5s07            &  6.58\pm 0.05 & ( 84.4\%)   &  6.50\pm 0.03 & ( 99.3\%)\\
o5s22            &  6.03\pm 0.02 & ( 98.2\%)   &  6.07\pm 0.04 & ( 54.3\%)\\
o5s24            & 13.54\pm 0.04 & ( 94.8\%)   & 13.70\pm 0.14 & ( 74.9\%)\\
o5t23            & 14.38\pm 0.08 & ( 74.0\%)   & 14.40\pm 0.09 & ( 41.3\%)\\
o5t32            & 11.34\pm 0.11 & ( 30.3\%)   & 11.30\pm 0.09 & ( 71.9\%)\\
o5t34            & 12.45\pm 0.04 & (100.0\%)   & 12.45\pm 0.04 & (100.0\%)\\
o5t49            & 12.66\pm 0.05 & ( 99.7\%)   & 12.72\pm 0.04 & (100.0\%)\\
\enddata%
\end{deluxetable}

\section{Discussion and conclusions}\label{sec:discussion}

\subsection{Implications for large surveys}\label{sec:implications}

Large variability can have an influence on whether or not an object is discovered in a magnitude limited survey.
Most large surveys use an automated moving object pipeline to find TNO candidates. 
For the Canada-France Ecliptic Plane Survey \citep{kavelaars09,petit11}, OSSOS and the \citet{alexandersen16} survey, TNOs were identified using three discovery images with one hour between each.
Non-stationary sources within three images of a field are linked together as a candidate TNO detection; various constraints on the rate, angle and linearity of motion are imposed, as well as a requirement of similar magnitudes, in order to limit the number of false candidates.
If a TNO's magnitude is highly variable, this might leave it undetected, either because it might be too faint to be detected in one of the three images or, if the constraints in the detection pipeline are too tight, because the pipeline might not consider the three points to be the same object due to having different magnitudes. 

\begin{figure}[tbp]
\includegraphics[width=1.0\textwidth]{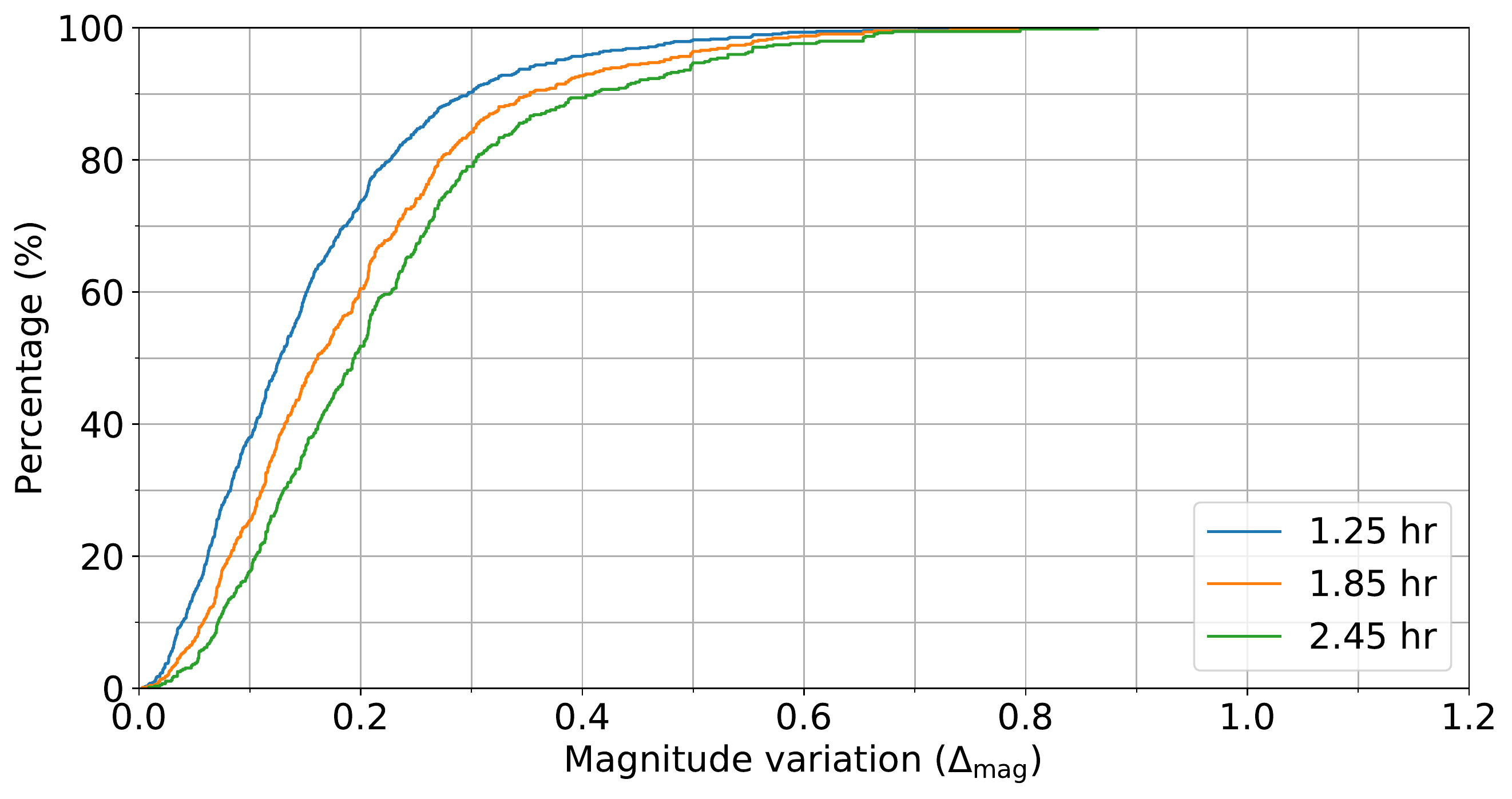}
\caption{\label{fig:cumulative_minmax}%
Cumulative distribution of the magnitude variation (maximum minus minimum, $\dmag$) within 1.25 hr (blue, top), 1.85 hr (orange, middle) and 2.45 hr (green, bottom) sections of light curves. 
}
\end{figure}

In our sample of 63 TNO light curves (most having observation on both nights), we find that only $6\pm1$ have a $\dmag<0.2$ magnitudes, while $31\pm2$ (just under half the sample) have a $\dmag>0.4$ magnitudes and $11\pm1$ have $\dmag>0.6$ magnitudes\footnote{The uncertainties quoted in this sentence on the number of objects come from resampling the light curve photometry 1000 times and calculating the values for each resampled set.}.
However, those variations are usually over many hours.
In order to evaluate the impact that light curve variation could have on shorter timescales, we measured the typical variation on 1--2 hour timescales. 
Since our images for these light curves are typically spaced 36 minutes apart, we calculated the $\dmag$ for all 1.25 hr (3 images), 1.85 hr (4 images) and 2.45 hr (5 images) subsets of our light curves. 
We find that in 1.25 hr, the median value of the $\dmag$ is 0.13 magnitudes and the standard deviation of the $\dmag$ distribution is 0.12 magnitudes. 
For 1.85 hr, the median and standard deviation are 0.16 and 0.13, respectively, while for 2.45~hr they are 0.19 and 0.14 magnitudes, respectively. 
\autoref{fig:cumulative_minmax} shows the $\dmag$ distribution within these three time spans. 
The moving object pipeline used for the \citet{alexandersen16} survey and for OSSOS only had the weak constraint that the faintest and brightest points within the three discovery images (spaced about 1 hour apart) had to have a flux-ratio of less than 4, which corresponds to a magnitude difference of about 1.5~mag. 
In our data from this work, we never see variability greater than 1.2~mag within a 2-hour window; it therefore seems highly unlikely that a real objects was rejected by the detection pipeline due to natural variability. 
Because the overall average variability seen within our data is much larger than the average variability measured within 1.25-2.45 hours, this demonstrates the perhaps obvious point that the average magnitude measured within a few hours is not an accurate measurement of the true average magnitude of the TNO. 
Discovery images from surveys like CFEPS, OSSOS and the \citet{alexandersen16} survey all depend on having measured the magnitude in three discovery images; the fact that these measured magnitudes are potentially several tenths of a magnitude from the true average magnitude should be included when performing survey simulation and model analysis. 

Variability is important to consider when measuring photometric colors of TNOs as well. 
Because most facilities do not have the ability to observe in multiple band-passes simultaneously, the images in different filters must be taken in sequence. 
If the object is faint and thus requires a long exposure time in some/each filter, its variability could cause an erroneous measurement of the color if the brightness varies significantly within the timespan of the observations. 
For accurate photometric colors, it is therefore important to account for light curve variability, except in cases of very short (thus nearly simultaneous) exposure sequences for bright objects.
This can be done by repeating observations of the filter that requires least exposure time several times during a sequence (for example $r$-$g$-$r$-$J$-$r$), as has become common practice, to get a rough idea of the light curve variation during the sequence. Alternatively, if some of the filters require enough exposure time to warrant multiple exposures, those exposures should be spread out in the sequence (for example the $r$-$g$-$J$-$g$-$r$ sequence used by \citet{bannister17}).
Of course, this relies on having fast filter changes and can therefore not be done with Hyper Suprime-Cam, which has a filter exchange time of about 30 minutes. 
It should be obvious that photometry acquired on different days cannot be used to calculate reliable colors unless the light curve is known, so TNO colors reported from surveys like the Hyper Suprime-Cam Subaru Strategic Program \citep[HSC SSP][]{aihara18, terai18} or the LSST survey should be seen as very rough color estimates with a likely uncertainty of $\sim0.2$ magnitude.

Expected to start science operations in 2022, LSST will provide an inventory of the trans-Neptunian region down to a 5-$\sigma$ limiting magnitude of approximately 24.6 in r and 24.8 in g \citep{ivezic08, LSST09}. 
LSST is expected to discover over 40,000 TNOs during its planned ten years of science operations \citep{LSST09}.
During this period, TNOs within the survey footprint will be imaged hundreds of times by LSST in multiple filters (\emph{ugrizy}), producing sparsely sampled light curves. Our HSC variability study overlaps the size and magnitude range that LSST will be sensitive to. 
Thus, the variability we observe can be taken as a minimum value for the variability LSST will see for its TNO sample.
Over a few hour baseline, 90\% of our HSC sample vary in brightness up to 0.4 magnitudes. 
This is more variability than shallower past wide-field Solar System surveys had to deal with. 
As detailed above, OSSOS, which our sample is derived from, had no significant brightness/variability constraints when identifying moving objects \citep{bannister18}.
This may be important in the development of the LSST moving object detection pipeline and for consideration when assessing the TNO detection completeness for LSST. 
Additionally, in the sparse light curves LSST will produce, objects with more than 0.5 magnitudes variation can potentially be flagged as extreme shapes in need of further higher-cadence follow-up observations. 

Finally, we note that objects in this size range do not have enough mass to ensure hydrostatic equilibrium, so the variability seen is therefore most likely caused by the objects being highly non-spherical. 
Given the correlation between absolute $H_r$ magnitude and variability shown above, it is unreasonable to expect that even smaller object can be modeled as roughly spherical shapes. 
A number of surveys have searched for serendipitous stellar occultations by small (diameters $\sim$1~km, roughly corresponding to $H_r\sim$16--19 depending on albedo) TNOs in order to measure the size distribution of such objects \citep{2016SPIE.9906E..5ML, 2013AJ....146...14Z, 2012ApJ...761..150S, 2009AJ....138..568B, 2008MNRAS.388L..44L, 2008AJ....135.1039B, 2003ApJ...594L..63R}. 
TNOs in this size range are in the Fresnel regime when measuring from the ground at visible wavelengths, and their occultation shadows thus exhibit significant diffraction features \citep{2007AJ....134.1596N, 1987AJ.....93.1549R}. 
Most survey teams use the approximation of spherical occulting objects in order to simplify the calculation of the occultation event parameters, but when non-spherical shapes are considered there may be significant effects on the resulting occultation shadows \citep{1987AJ.....93.1549R, joel}. 
Obviously, we did not measure light curves of such small objects in this survey; however, based on our observations of larger (8>$H_r$>10) TNOs, we propose that these surveys need to consider the expectation of non-spherical objects when defining event detection criteria, characterizing any occultation events or estimating any survey detection efficiencies.

\subsection{Contact binaries}

To date, in the trans-Neptunian belt, only 2001~QG$_{298}$ has been confirmed as a contact binary \citep{sheppardjewitt04, lacerda14a}.
This confirmation is based on the large light curve amplitude $\dmag\geq0.9$~mag, and the V-/U-shape at the minimum/maximum of brightness. 
However, such a large amplitude is only observed if the two components are seen equator-on (or almost equator-on). 
If the two components are not aligned with the viewer's line of sight, the amplitude will be smaller \citep[simple geometry,][]{lacerda11,lacerda14a,thirouin17a,thirouin17b,thirouinsheppard18}. 

Our sparse light curves reveals no obvious contact binaries, although do contain candidates that should be observed further. 
None of the light curves reported here present a definite full peak-to-peak amplitude larger than 0.9~mag.
05s49, uo5s77 and 05s57 have a $\dmag$ around 1~mag, but the light curves are too noisy for drawing conclusions. 
Several objects like 2003 SP$_{317}$, o5t30, 2013 UN$_{15}$ and 2013 SM$_{100}$ display a $\dmag>0.5$~mag. 
Unfortunately, because we do not have a full maximum and minimum showing a clear U- and V-shape, it is difficult to classify them as likely contact binaries. 
More observations are required in order to confirm whether these objects are contact binaries or simply highly elongated bodies. 


\subsection{Conclusions}

We have obtained a large sample of sparse partial light curves 65 TNOs, most of which are small ($H_r$>7) and therefore fainter than have typically been studied before, with up to 22 observations in acceptable conditions over two nights. 
While the variability for many of these objects might only be a lower limit due to not having observed the full rotation of the object, we can draw some tentative conclusions. 
As seen by \citet{benecchisheppard13}, we confirm that the amplitude of the light curve correlates with the absolute magnitude. 
However, in contrast, we do not see a statistically significant correlation with inclination. 

We find that a large fraction of our observed TNOs have large brightness variations, even on 1-2 hour timescales. 
31 objects (just under half the sample) have a $\dmag>0.4$~mag over the course of our observations, while the median variability within a 1.85~hour timespan was 0.16~mag. 
TNO variability must be accounted for when simulating/modeling discovery surveys like OSSOS and LSST, as well as when observing non-simultaneous TNO colors.

In this pilot study we did not find any significant difference in amplitude distribution between different dynamical classes. 
However, our sample included very few objects in the scattering and detached classes. 
Future work should therefore target more of these objects in order to have similar sample sizes for each population. 

We were able to estimate periods for 12 of the 65 TNOs.
However, these periods are not certain due to aliasing.
In order to estimate periods for a larger fraction of the sample and reduce aliasing effects, a longer baseline of observations is needed.


\section*{Acknowledgements}

This paper is based on data collected at the Subaru Telescope and retrieved from the Hyper Suprime-Cam data archive system, which is operated by Subaru Telescope and Astronomy Data Center at National Astronomical Observatory of Japan. 
The access to Subaru telescope from Taiwan is supported by the Academia Sinica Institute of Astronomy and Astrophysics.
The authors wish to recognize and acknowledge the very significant cultural role and reverence that the summit of Maunakea has always had within the indigenous Hawaiian community. We are most fortunate to have the opportunity to conduct observations from this mountain.
We would also like to acknowledge the maintenance, cleaning, administrative and support staff at academic and telescope facilities, whose labor maintains the spaces where astrophysical inquiry can flourish.

This work is based on TNO discoveries obtained with MegaPrime/MegaCam, a joint project of the Canada France Hawaii Telescope (CFHT) and CEA/DAPNIA, at CFHT which is operated by the National Research Council (NRC) of Canada, the Institute National des Sciences de l'Universe of the Centre National de la Recherche Scientifique (CNRS) of France, and the University of Hawaii. 
A portion of the access to the CFHT was made possible by the Academia Sinica Institute of Astronomy and Astrophysics, Taiwan. 
This research used the facilities of the Canadian Astronomy Data Centre operated by the National Research Council of Canada with the support of the Canadian Space Agency. 

MES was supported by Gemini Observatory, which is operated by the Association of Universities for Research in Astronomy, Inc., on behalf of the international Gemini partnership of Argentina, Brazil, Canada, Chile, and the United States of America. 
MB acknowledges support from UK STFC grant ST/L000709/1.
KV acknowledges support from NASA grants NNX14AG93G and NNX15AH59G. 
SBD acknowledges support by the National Aeronautics and Space Administration under Grant/Contract/Agreement No. NNX15AE04G issued through the SSO Planetary Astronomy Program.

This paper makes use of software developed for the Large Synoptic Survey Telescope. We thank the LSST Project for making their code available as free software at \url{http://dm.lsst.org}. 
This research has made use of NASA's Astrophysics Data System Bibliographic Services. 
This research made use of SciPy \citep{jones01}, NumPy \citep{vanderwalt11}, matplotlib \citep[a Python library for publication quality graphics][]{hunter07} and Astropy \citep[a community-developed core Python package for Astronomy][]{astropy1, astropy2}. 
This research made use of ds9, a tool for data visualization supported by the Chandra X-ray Science Center (CXC) and the High Energy Astrophysics Science Archive Center (HEASARC) with support from the JWST Mission office at the Space Telescope Science Institute for 3D visualization. 
IRAF is distributed by the National Optical Astronomy Observatory, which is operated by the Association of Universities for Research in Astronomy (AURA) under cooperative agreement with the National Science Foundation \citep{tody93}. 

The Pan-STARRS1 Surveys (PS1) and the PS1 public science archive have been made possible through contributions by the Institute for Astronomy, the University of Hawaii, the Pan-STARRS Project Office, the Max-Planck Society and its participating institutes, the Max Planck Institute for Astronomy, Heidelberg and the Max Planck Institute for Extraterrestrial Physics, Garching, The Johns Hopkins University, Durham University, the University of Edinburgh, the Queen's University Belfast, the Harvard-Smithsonian Center for Astrophysics, the Las Cumbres Observatory Global Telescope Network Incorporated, the National Central University of Taiwan, the Space Telescope Science Institute, the National Aeronautics and Space Administration under Grant No. NNX08AR22G issued through the Planetary Science Division of the NASA Science Mission Directorate, the National Science Foundation Grant No. AST-1238877, the University of Maryland, Eotvos Lorand University (ELTE), the Los Alamos National Laboratory, and the Gordon and Betty Moore Foundation.

The acknowledgements were compiled in part using the Astronomy Acknowledgement Generator. 

\vspace{5mm}
\facilities{Subaru (HSC), PS1}

\software{Python \citep{vanrossumdeboer91}, Matplotlib \citep{hunter07}, NumPy \citep{vanderwalt11}, SciPy \citep{jones01}, TRIPPy \citep{fraser16}, Ureka \citep{hirst14}, IRAF \citep{tody93}, Astropy \citep{astropy1,astropy2}, SExtractor \citep{bertin96}, ds9 \cite{joye03}}


\bibliography{lc_ossos_hsc}


\appendix

\subsection{Supplementary materials}

\begin{figure}[b]
\plotone{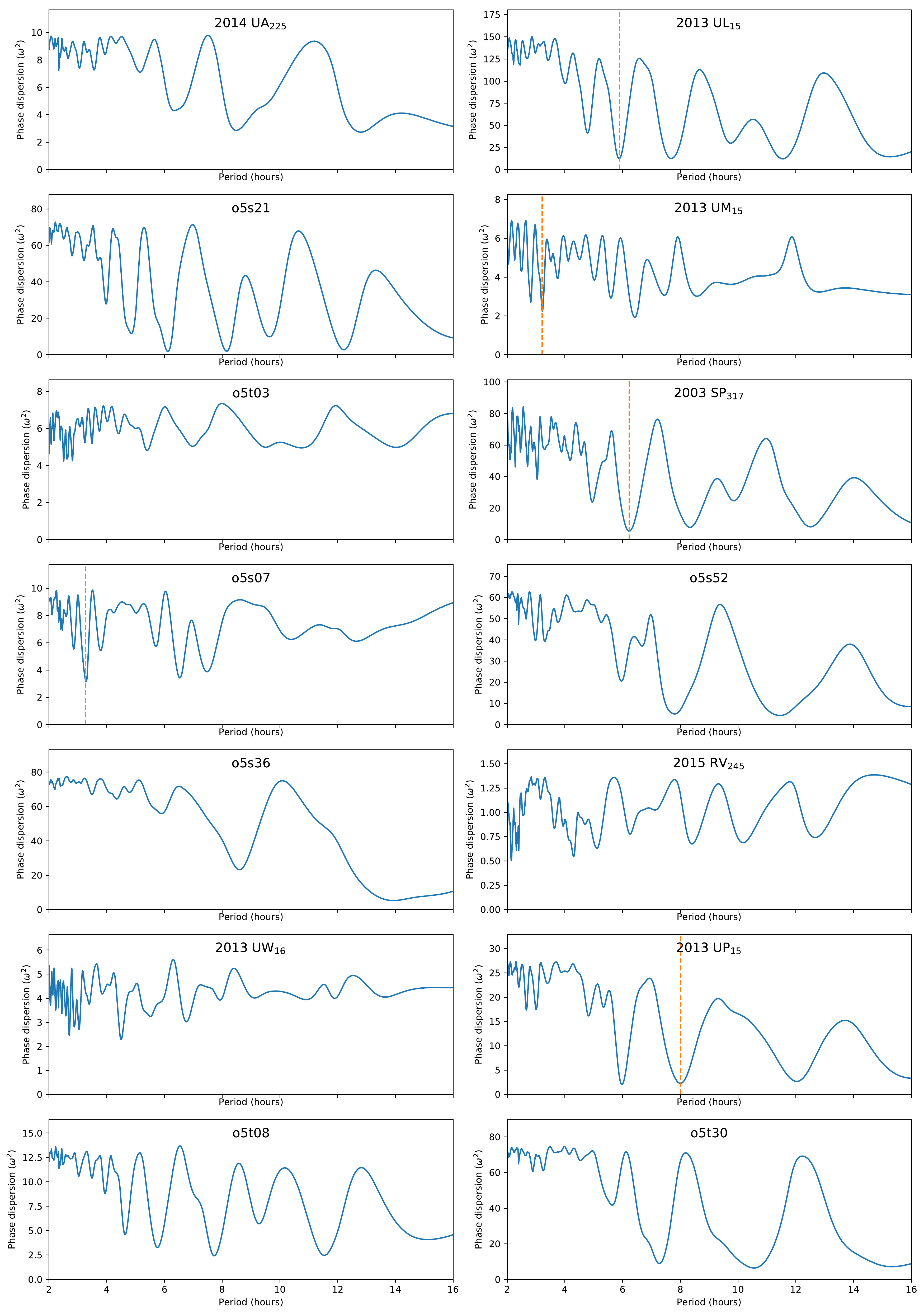}
\caption{\label{fig:pdm0}%
PDM plot from our period analysis. Vertical dashed orange lines show our favored single-peak period for the objects listed in \autoref{tab:periods} and shown in \autoref{fig:folded}. 
Objects 0--13, ordered by average magnitude. 
}
\end{figure}

\begin{figure}[tbp]
\plotone{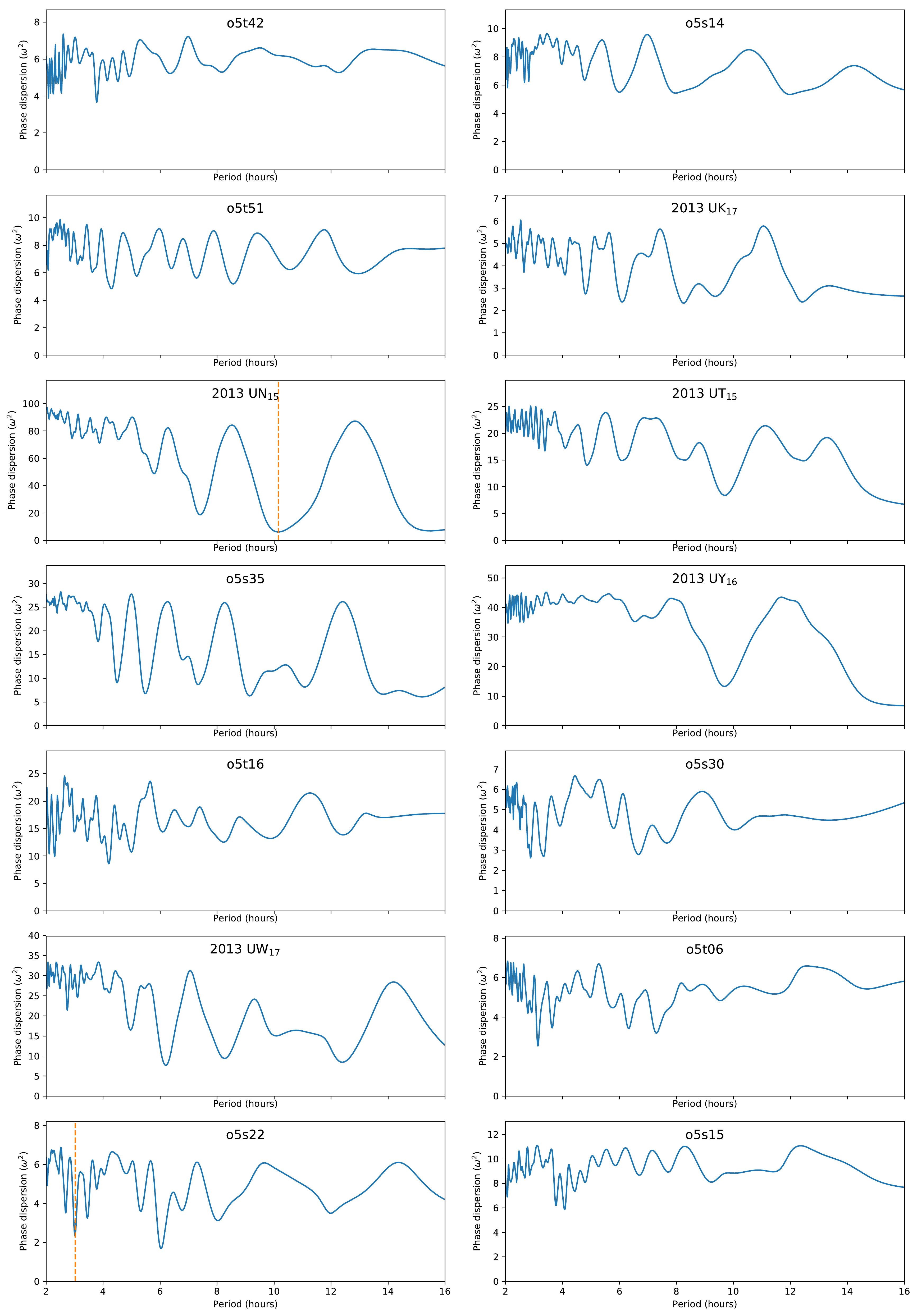}
\caption{\label{fig:pdm1}%
PDM plot from our period analysis. Vertical dashed orange lines show our favored single-peak period for the objects listed in \autoref{tab:periods} and shown in \autoref{fig:folded}.
Objects 14--27, ordered by average magnitude. 
}
\end{figure}

\begin{figure}[tbp]
\plotone{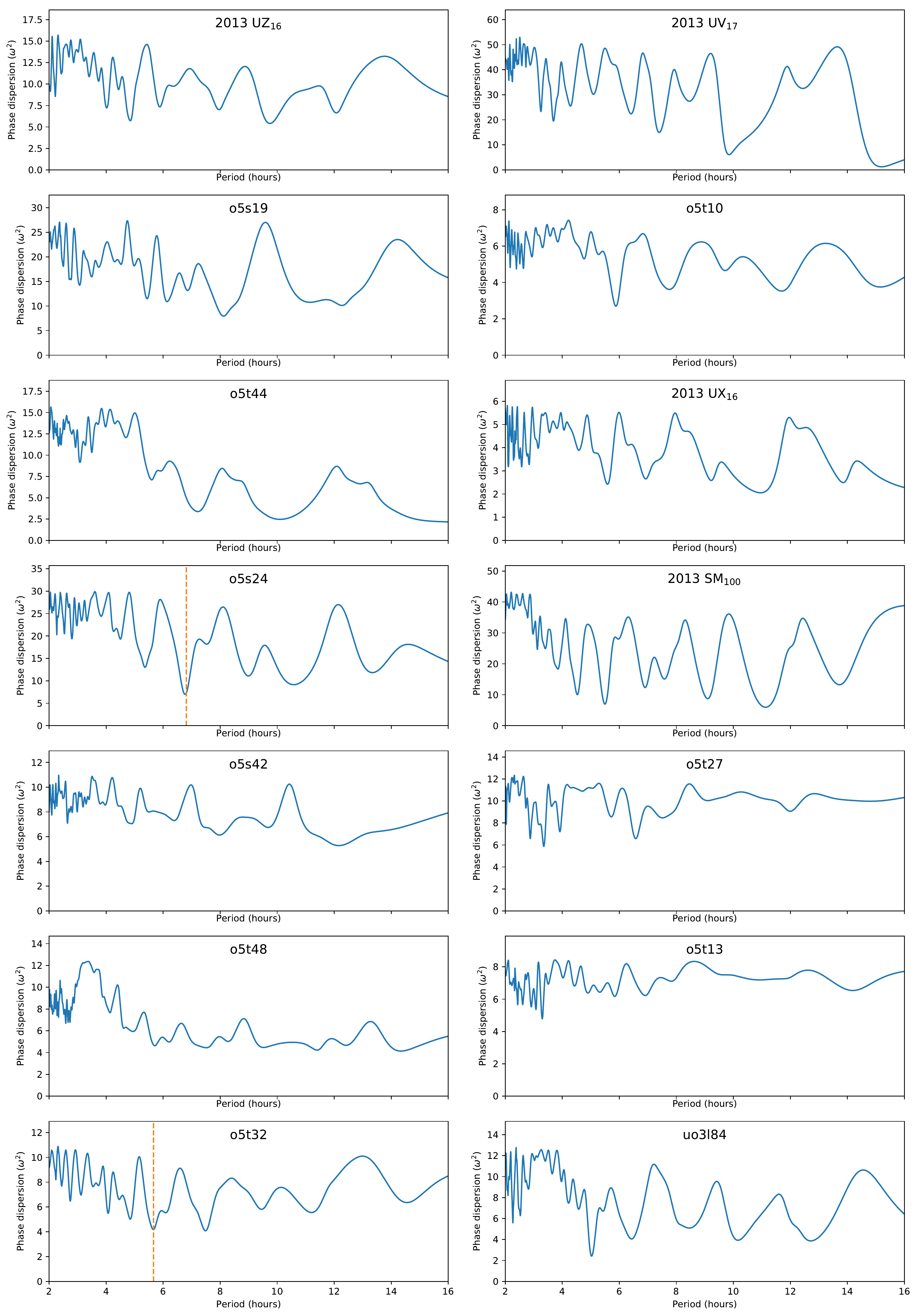}
\caption{\label{fig:pdm2}%
PDM plot from our period analysis. Vertical dashed orange lines show our favored single-peak period for the objects listed in \autoref{tab:periods} and shown in \autoref{fig:folded}.
Objects 28--41, ordered by average magnitude. 
}
\end{figure}

\begin{figure}[tbp]
\plotone{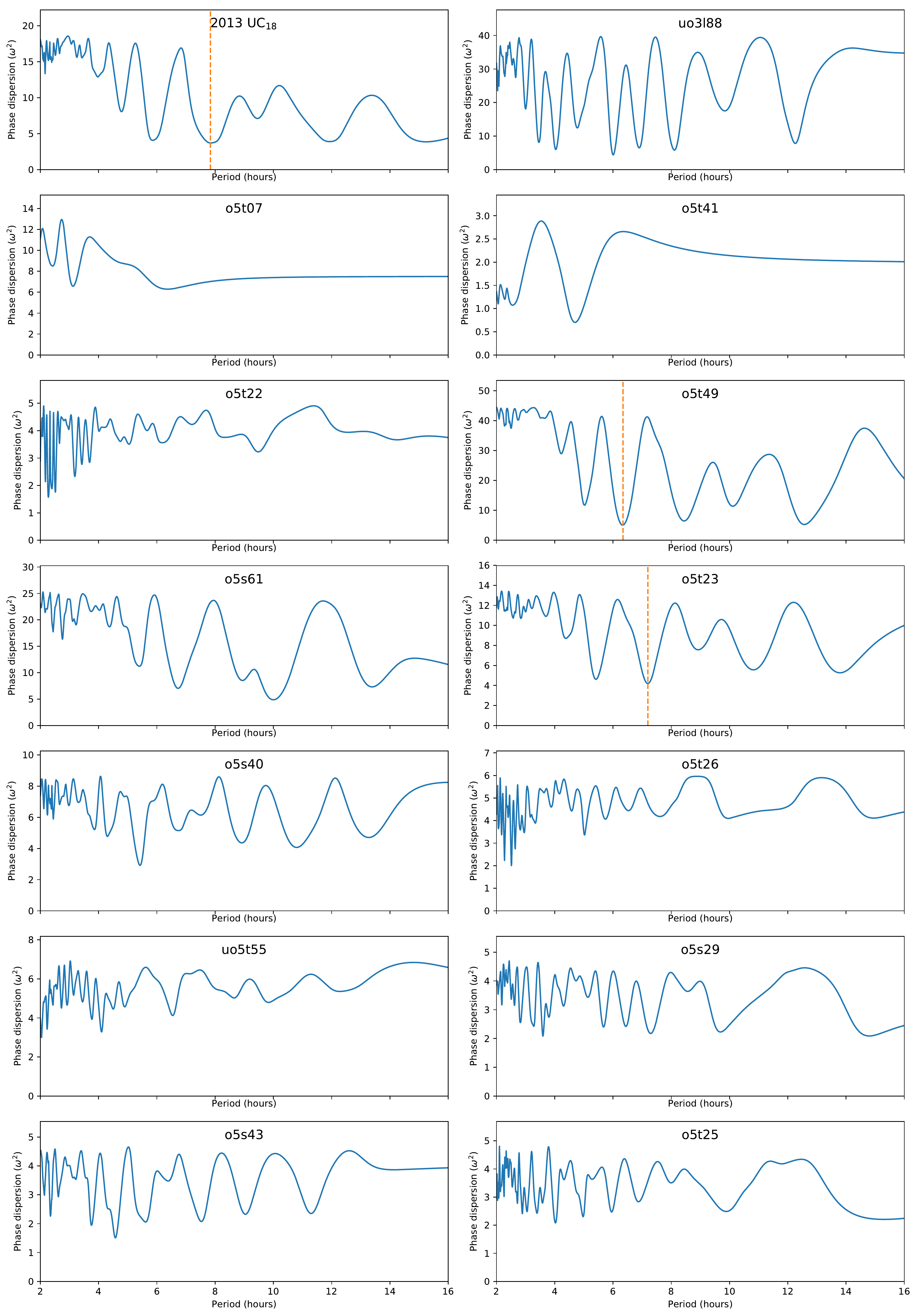}
\caption{\label{fig:pdm3}%
PDM plot from our period analysis. Vertical dashed orange lines show our favored single-peak period for the objects listed in \autoref{tab:periods} and shown in \autoref{fig:folded}.
Objects 42--55, ordered by average magnitude. 
}
\end{figure}

\begin{figure}[tbp]
\plotone{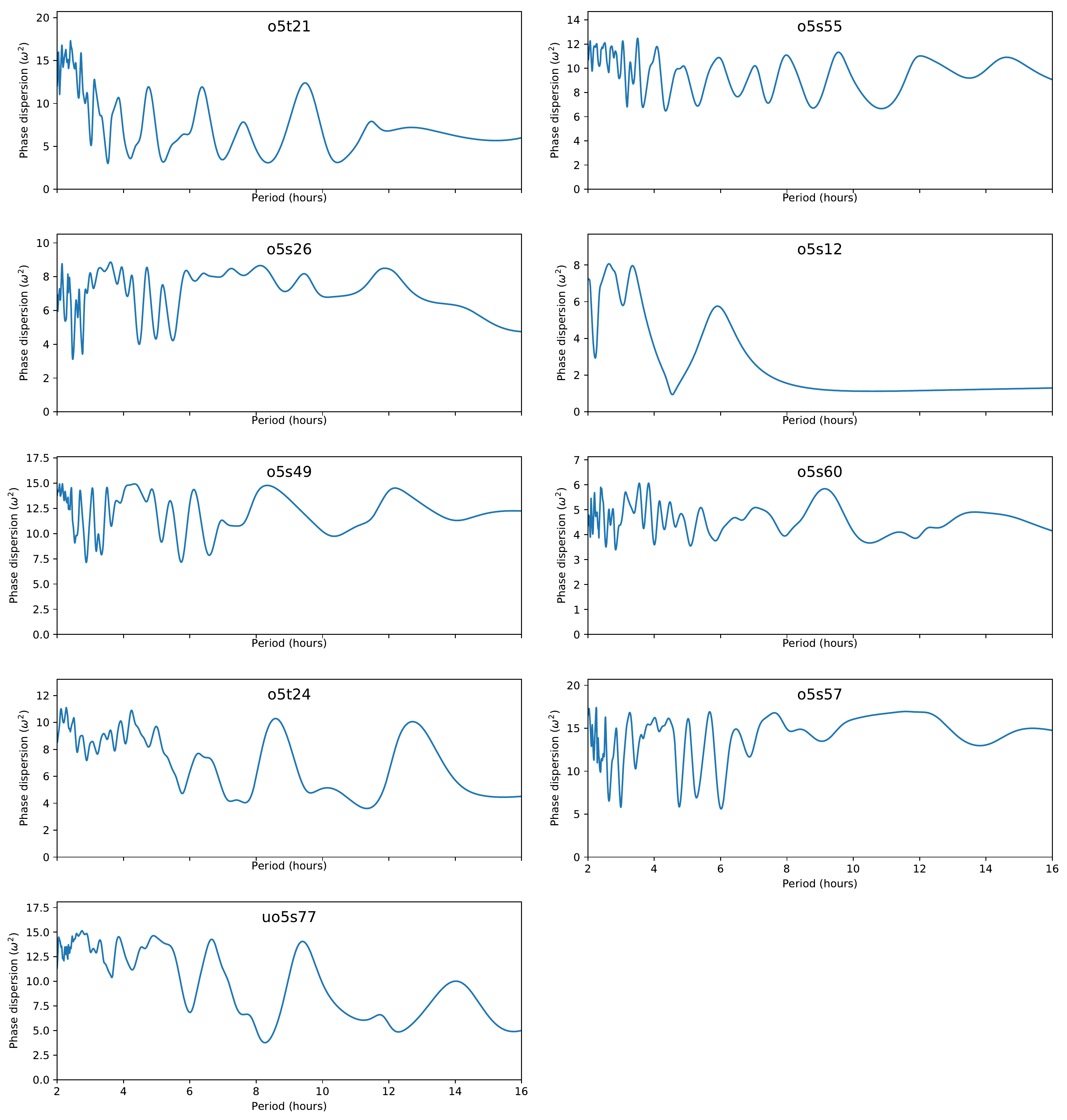}
\caption{\label{fig:pdm4}%
PDM plot from our period analysis. 
Objects 56--64, ordered by average magnitude. 
}
\end{figure}

\clearpage
\clearpage
\clearpage

\startlongtable
\begin{deluxetable}{CCC}
\tabletypesize{\scriptsize}
\tablecaption{\label{tab:photometry_long} Photometry from HSC data.
$sigma_{Rand}$ is the random uncertainty of the $m_r$ magnitude measurement.
$Zp$ is the zero-point used. 
$sigma_{Sys}$ is the systematic uncertainty on the zero-point (and thus also on the measurement).
}
\tablehead{%
\colhead{MJD$-57620$} & \colhead{$m_r$ $\pm$ $\sigma_{Rand}$} & \colhead{$Zp$ $\pm$ $\sigma_{Sys}$}
}
\startdata%
          & \textrm{o3l19}     &                     \\
6.3913125 & 24.09   \pm 0.03   & 26.995  \pm 0.005   \\
6.4154337 & 24.21   \pm 0.02   & 27.012  \pm 0.005   \\
6.4393438 & 24.45   \pm 0.02   & 27.022  \pm 0.005   \\
6.4634385 & 24.69   \pm 0.04   & 27.058  \pm 0.005   \\
6.4876346 & 24.78   \pm 0.03   & 27.047  \pm 0.005   \\
6.5123402 & 24.56   \pm 0.03   & 27.039  \pm 0.005   \\
6.5375195 & 24.41   \pm 0.03   & 27.043  \pm 0.005   \\
\enddata%
\end{deluxetable}

\end{document}